\documentclass[12pt]{article}
\usepackage[left=2cm, right=2cm, top=2cm, bottom=2cm]{geometry}
\usepackage{amsmath}
\usepackage{SIunits}
\usepackage{graphicx}
\usepackage{placeins}       
\usepackage{pifont}         

\begin{document}

\textbf{Are Small Polarons Always Detrimental to Transparent Conducting Oxides ?}

\textit{Guillaume Brunin, Gian-Marco Rignanese, and Geoffroy Hautier*}

G. Brunin, Prof. G.-M. Rignanese, Prof. G. Hautier

UCLouvain, Institut de la Mati\`ere Condens\'ee et des Nanosciences (IMCN), Chemin des \'Etoiles~8, Louvain-la-Neuve 1348, Belgium

Correspondence and requests for materials should be addressed to G.H. 

(email:~geoffroy.hautier@uclouvain.be)

\vspace{1cm}


Transparent conducting oxides (TCOs) are essential to many technologies including solar cells and transparent electronics. The search for high performance n- or p-type TCOs has mainly focused on materials offering transport through band carriers instead of small polarons. In this work, we break this paradigm and demonstrate using well-known physical models that, in certain circumstances, TCOs exhibiting transport by small polarons offer a better combination of transparency and conductivity than materials conducting through band transport. We link this surprising finding to the fundamentally different physics of optical absorption for band and polaronic carriers. Our work rationalizes the good performances of recently emerging small-polaronic Cr-based p-type TCOs such as Sr-doped LaCrO$_3$ and outlines design principles for the development of high-performance TCOs based on transport by small polarons. This opens new avenues for the discovery of high-performance TCOs especially p-type.
 
\newpage


Transparent conducting oxides (TCOs) are exceptional materials combining the antagonistic properties of optical transparency and electrical conductivity~\cite{ellmer2012past,kehoe2016assessing}.
The transparency of TCOs is achieved using wide band gap oxides ($E_g > 3$~eV) and they are made highly conductive by the introduction of excess electrons (n-type) or holes (p-type) through intrinsic or extrinsic doping. 
High performance TCOs are essential to many modern technologies including solar cells, optoelectronic devices, touch screens and light emitting diodes~\cite{kehoe2016assessing,minami2005transparent,hautier2013identification}.
The best n-type TCOs, based on doped In$_2$O$_3$, ZnO or SnO$_2$, show conductivities on the order of $10^{4}$~S~cm$^{-1}$ and transmittances well above 80\%~\cite{minami2005transparent,Castaneda2011,tate2002p,zhang2016p,edwards2004basic}.
On the other hand, p-type TCOs such as ZnRh$_2$O$_4$, CuAlO$_2$, SrCu$_2$O$_2$ show much poorer performances. 
Their conductivities reach at most 10~S~cm$^{-1}$ for transparencies between 50 and 80~\% (for films of a few hundred nm thicknesses)~\cite{hautier2013identification,zhang2016p,Dekkers2007}.
This lack of high-performance p-type TCOs limits many applications in transparent electronics or new solar cell architectures~\cite{zhang2016p}.

Very recently, Cr-based oxides have shown attractive performances as p-type TCOs. 
Materials such as Cu-deficient CuCrO$_2$, Mg-doped Cr$_2$O$_3$ and more recently Sr-doped LaCrO$_3$ have shown among the best transparencies and conductivities for p-type oxides~\cite{zhang2015perovskite,zhang2015hole,farrell2016synthesis,arca2011magnesium}.
These Cr-based oxides exhibit a thermally activated mobility because charge transport occurs through small-polaron hopping. 
A small polaron consists of a charge carrier localized on a single atomic site due to the interactions with the surrounding ions. 
Small-polaron transport has usually been considered detrimental for TCO applications~\cite{kehoe2016assessing,zhang2016p,farrell2015conducting,varley2012role}.
Indeed, the low mobilities of small polarons ($\mu \ll 1$~cm$^2$V$^{-1}$s$^{-1}$) compared to band carriers ($\mu \sim 0.1-100$~cm$^2$V$^{-1}$s$^{-1}$)\cite{bosman1970small} are believed to prevent them to achieve high performances as TCOs. And even for applications in which conductivity (and not mobility) is the most important property, this low mobility needs to be counterbalanced by a higher carrier concentration which is traditionally correlated with the degradation of transparency~\cite{bellingham1992intrinsic}.

These considerations have led the TCO community to advocate against the use of materials based on small-polaron transport. Naturally, this affected the search and design of new TCOs (including those computationally-driven) with the avoidance of small-polaron formation often put forward as a criteria for a high performance TCO~\cite{peng2013li}.
The apparent contradiction between this criteria and the encouraging performances of Cr-based oxides has not been addressed so far in the literature. 
A clear rationalization of the true inconvenients and potentially overlooked benefits of TCOs based on small-polaron transport is therefore greatly needed.

Here, we use well-established physical models describing how transparency and conductivity vary with materials parameters (effective mass, carrier concentration,...) to compare the performances of ideal band and small-polaron TCOs. 
We show that TCOs relying on small-polaron transport (SP-TCOs) can outperform those relying on band transport (B-TCOs) especially when the latter exhibit a high effective mass (which is often the case for p-type oxides). 
The physical origin of this surprising result lies in the very different change of the optical absorption of B- and SP-TCOs as the carrier concentration increases. 
While B-TCOs see their optical transparency drop when high carrier concentrations are reached, SP-TCOs can remain highly transparent. 
Our results rationalize the good performances of Cr-based p-type polaronic oxides and motivate the TCO community to revise its avoidance for small-polaron materials. 
We also provide a series of design principles for the search of novel SP-TCOs. 
This opens an entirely new avenue towards the search and development of high performance TCOs expecially p-type.

\section*{Results}

\subsection*{Comparison of performances for B- and SP-TCOs}

\begin{figure}[!h]
	\center
		{\includegraphics[clip,trim=0.6cm 0cm 0.1cm 0.2cm,width=.75\columnwidth]{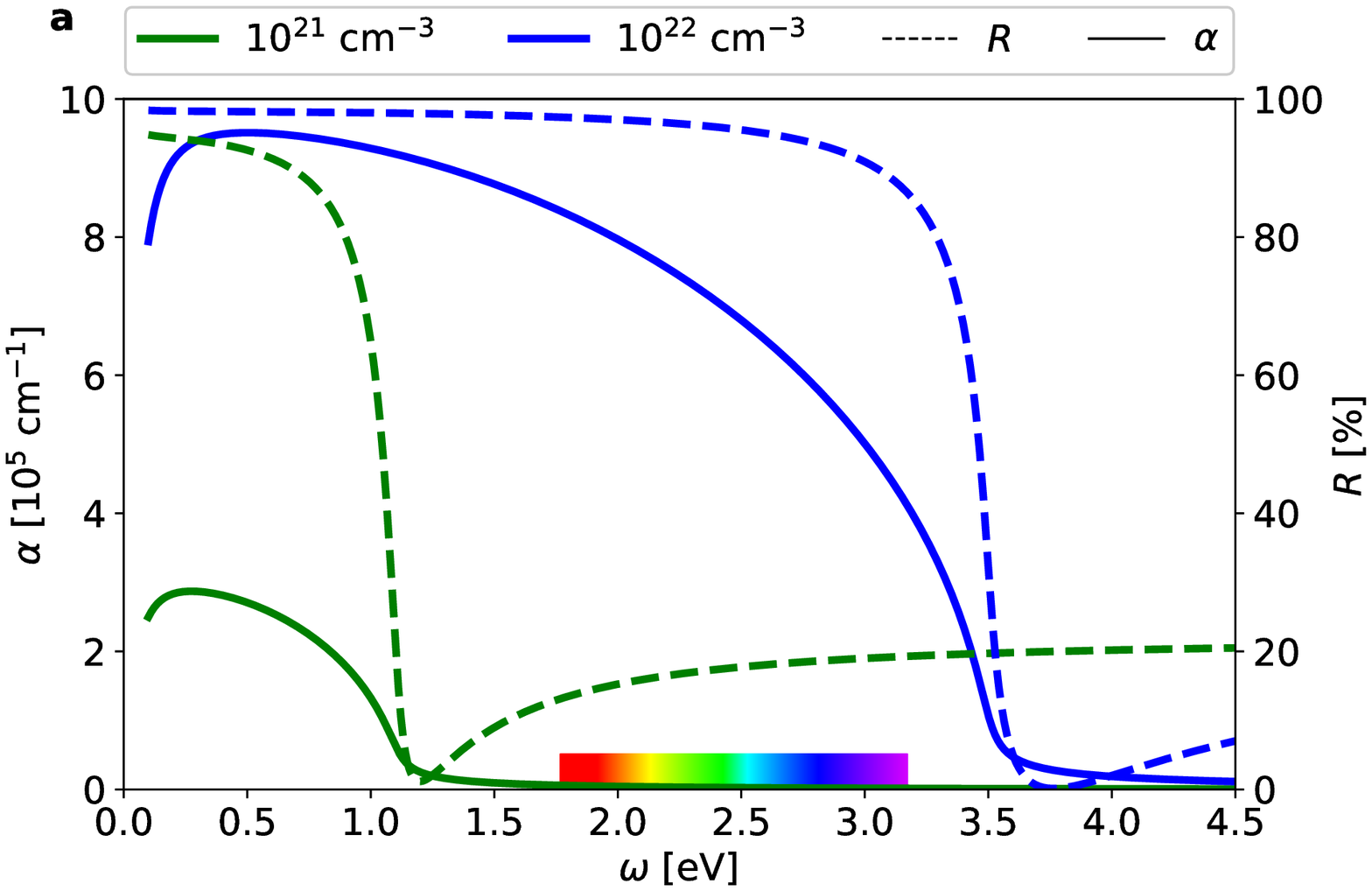}}
		
		\vspace{3mm}
		{\includegraphics[clip,trim=0.6cm 0cm 0.1cm 0.1cm,width=.75\columnwidth]{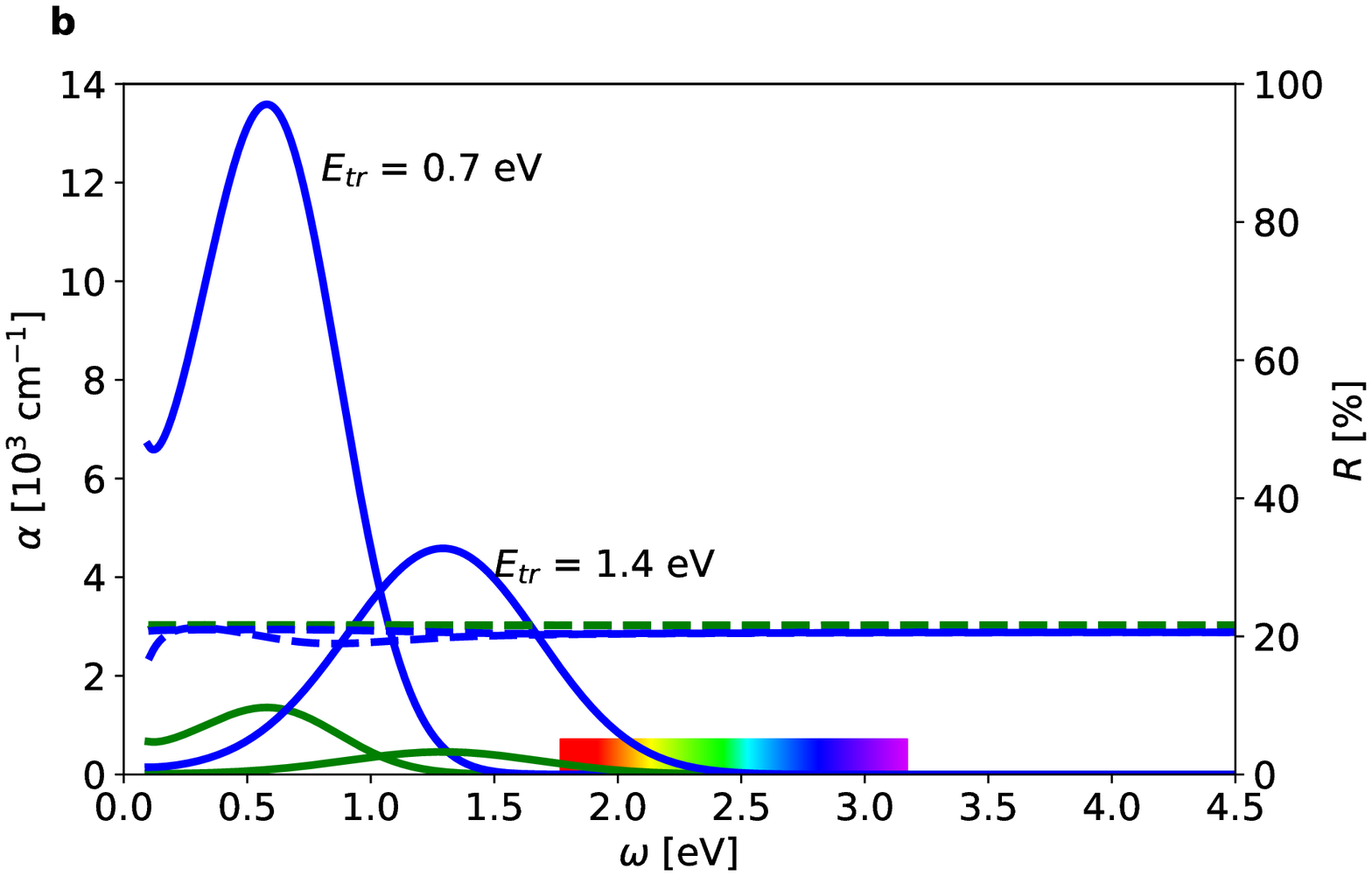}}
	\caption{ Absorption coefficient $\alpha$ and reflectivity $R$ for (\textbf{a}) B-TCOs with an effective mass $m^* = 0.15 m_0$ and mobility $\mu = 90$~cm$^2$V$^{-1}$s$^{-1}$ and (\textbf{b}) SP-TCOs for two transition energies. Different carrier concentrations are shown $10^{21}$~cm$^{-3}$ (in green) and $10^{22}$~cm$^{-3}$ (in blue).}
	\label{fig:Polaron_vs_Drude}
	\end{figure}

\begin{figure}[!h]
	\center
	\includegraphics[clip,trim=0.3cm 0.4cm 1.6cm 1.9cm,width=.8\columnwidth]{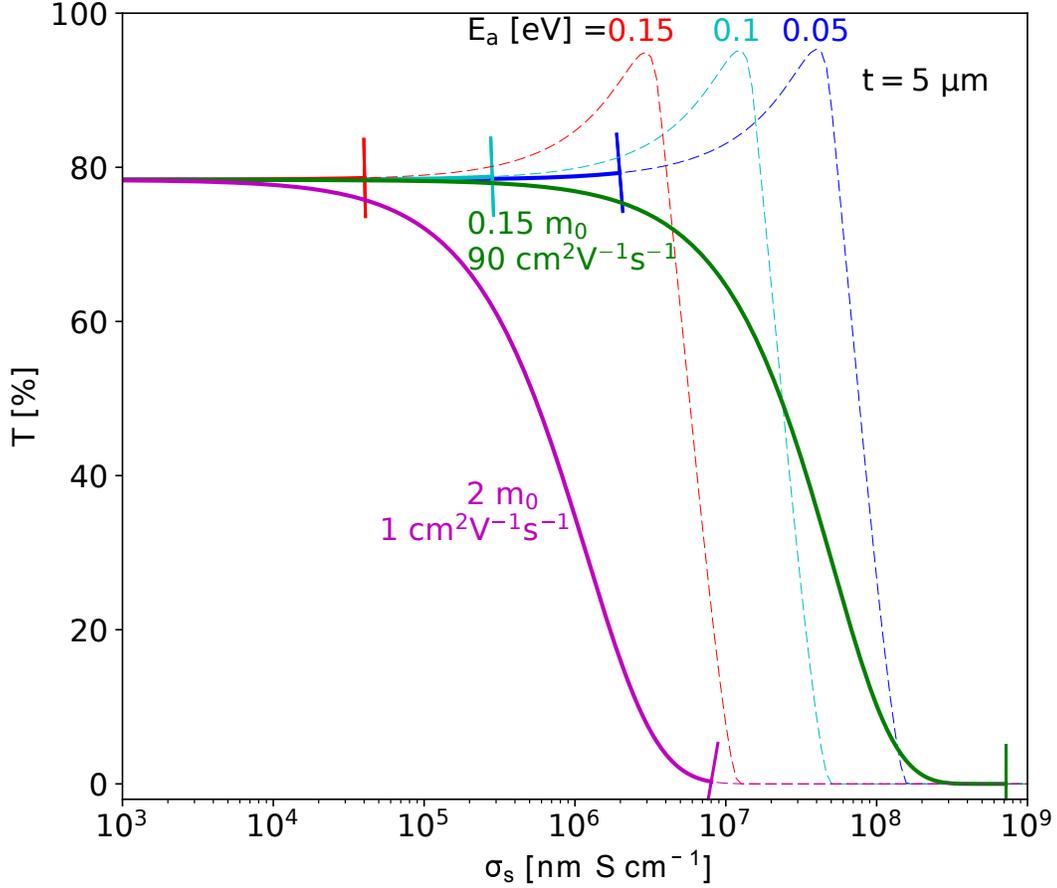}
	\caption{Average transmittance $T$ of visible light versus sheet conductivity $\sigma_s$ for low effective mass/high mobility (green) and high effective mass/low mobility (purple) B-TCOs, and SP-TCOs (blue, cyan and red) with various transport activation energies $E_a$. We used a thickness $t = $~\unit{5}{\micro\meter}. Solid (resp. dashed) lines correspond to carrier concentrations lower (resp. higher) than $10^{22}$~cm$^{-3}$.}
	\label{fig:Transmittance_Conductivity}
	\end{figure}

The optical response of highly doped B-TCOs typically follows the Drude model which assumes an electron gas in a potential imposed by the nuclei. 
Figure~\ref{fig:Polaron_vs_Drude}a plots the reflectivity $R$ and absorption coefficient $\alpha$ as functions of the energy $\omega$ of the light at different doping levels and for different material parameters (effective mass, mobilities) for the Drude model (see Supplementary Information, Section 1 and 2 for the full derivation). 
Below a certain threshold for the energy of the light, both the reflectivity $R$ and the absorption coefficient $\alpha$ are large due to collective oscillations of the carrier gas. 
In this regime, the highly doped B-TCO behaves as a metal. 
The threshold in energy below which collective oscillations of the carriers happen corresponds to the so-called plasma frequency $\omega_p$:
\begin{equation}
\omega_p^2 = \frac{Ne^2}{\varepsilon_r \varepsilon_0 m^*}
\end{equation}
where $N$ is the carrier concentration, $e$ is the elementary charge, $\varepsilon_r$ is the high-frequency limit of the real part of the dielectric function, $\varepsilon_0$ is the vacuum permittivity and $m^*$ is the effective mass of the band carriers.
When the carrier concentration in a B-TCO increases, the plasma frequency shifts towards higher energies and from the infra-red to the visible spectrum. 
If the carrier concentration is too large, most visible light is reflected by the material while the rest is absorbed. 
This is clearly illustrated in Fig.~\ref{fig:Polaron_vs_Drude}a where a material transparent in the visible spectrum at $10^{21}$~cm$^{-3}$ (in green) reflects and absorbs most of the visible light when doped to $10^{22}$~cm$^{-3}$ (in blue). 
This phenomenon sets a limit to the doping level and therefore to the conductivity reachable in a B-TCO. As a result, a compromise needs to be found between maximizing the conductivity and the transmittance~\cite{bellingham1992intrinsic}.
The transparency-conductivity compromise can be directly probed by plotting transmittance versus the sheet conductivity (i.e., the product of the conductivity and the thickness $t$ of the material)~\cite{fleischer2017quantifying}. 
Figure~\ref{fig:Transmittance_Conductivity} reports the transmittance versus the sheet conductivity for B-TCOs with different materials properties: in green for low effective mass/high mobility (best values for n-type B-TCOs \cite{ellmer2012past,zhang2016p}) and in purple for high effective mass/low mobility (typical good values for p-type B-TCOs \cite{hautier2013identification,zhang2016p}) materials.
We have also considered other typical n-type TCOs parameters (mobility $\mu = 50$ cm$^2$V$^{-1}$s$^{-1}$ and effective mass $m^* = 0.3 m_0$ \cite{ellmer2012past}), and the transmittance versus sheet conductivity curve is similar to the one corresponding to the best parameters (larger mobility). The difference between the two is that a larger carrier concentration is required when the mobility is lower to reach the same sheet conductivity, and a larger effective mass decreases the plasma frequency, allowing for larger carrier concentrations before the loss of transmittance.
Solid lines in Fig.~\ref{fig:Transmittance_Conductivity} indicate carrier concentrations lower than $10^{22}$~cm$^{-3}$ while higher carrier concentrations are shown by dashed lines (as it is unlikely that higher carrier concentrations could be reached).
For both low and high effective mass materials, increasing the sheet conductivity through higher carrier concentration leads at some point to a strong degradation of the transmittance. 
This degradation comes directly from the movement of the plasma edge towards higher frequencies. 
The low and high effective mass materials behave quantitatively differently: a higher sheet conductivity and transmittance can be reached for high mobility materials. 
Hence, justifying the interest for high mobility/low effective mass TCOs as they offer larger transmittance and sheet conductivity when band transport is involved~\cite{zhang2016p}.

\FloatBarrier

The optical response of SP-TCOs is extremely different from the one of B-TCOs. 
In the former, the carriers are trapped and cannot oscillate collectively. 
Optical transitions (additionally to the band gap interband transition) are however present because of the small polarons. 
These can be excited to either a delocalized (free) state or to a small-polaron state at an adjacent site (see Supplementary Information, Section 3). 
The physics of these processes has been studied and the absorption coefficient of a material containing small polarons is given by~\cite{emin1993optical}
\begin{equation}
\alpha(\omega) = K_1 \frac{N}{\omega} \exp \left( \frac{-(E_{tr} - \hbar\omega)^2}{\Delta^2} \right) \label{eq:absorption}
\end{equation}
with $N$ the small-polaron concentration, $E_{tr}$ the energy of the small-polaron transition, $\omega$ the frequency of the incoming light and $K_1$ a constant defined in Supplementary Information, Section 3. 
The constant $K_1$ depends on the phonon and electronic properties as well as on the structure of the material but it is not expected to change significantly, especially as it is only a prefactor to the dominant exponential term.
$\Delta$ is the phonon broadening of the ground state.
The absorption coefficient of an SP-TCO is plotted in Fig.~\ref{fig:Polaron_vs_Drude}b for two arbitrary transition energies (0.7 and 1.4~eV) and for different small-polaron concentrations (in green and blue). 
This figure also reports the reflectivity of SP-TCOs (see Supplementary Information, Section 3). The latter is dominated by the polarization of ion cores which leads to a constant reflectivity in the whole range and is characterized by $\varepsilon_r$.
The effect of small polarons, which leads to deviations with respect to this constant and depends on their transition energy and their concentration, is not significant for visible transparency.
The curves corresponding to a large set of transition energies and concentrations are available in Supplementary Figs.~4 to 8.

Figure~\ref{fig:Polaron_vs_Drude}b shows that as the small-polaron concentration increases, only the absorption peak height is modified and not its position. 
This behavior is remarkably different from the one of B-TCOs in which the absorption edge shifts to higher energies when the carrier concentration increases (see Fig.~\ref{fig:Polaron_vs_Drude}a). 
If the small-polaron transition energy is small enough, the absorption peak is not significant in the visible spectrum and the SP-TCO remains transparent even when very high carrier concentrations are reached ($10^{22}$~cm$^{-3}$).

The physics of the transport in SP-TCOs is also very different from B-TCOs. 
The mobility $\mu$ of small polarons is thermally activated through an activation energy $E_a$ needed for the polaron to hop from site to site~\cite{devreese1996polarons}:
\begin{equation}
\mu = \frac{K_2}{k_B T} \exp\left(\frac{-E_a}{k_BT}\right) \label{eq:mobility}
\end{equation}
with $K_2$ a constant defined in Supplementary Information, Section 3, $T$ the temperature, and $k_B$ the Boltzmann constant. 
Again, the constant $K_2$ depends on the phonon properties and on the structure of the material but it is not expected to change significantly, as compared to the exponential factor.
The mobility directly impacts the conductivity and is much lower than 1~cm$^2$V$^{-1}$s$^{-1}$ when small polarons are involved.
As we did for B-TCOs, we can combine the conductivity of small polarons with their optical response and analyze how the transmittance of a polaronic thin film changes with sheet conductivity. 
For SP-TCOs, there are two main parameters: the transition energy ($E_{tr}$) which determines the optical properties and the activation energy ($E_a$) which sets the electrical properties. 
In the commonly used Marcus theory these two quantities are actually linked  through $E_a = \frac{E_{tr}}{4}$ \cite{mildner2015temperature,deskins2007electron}. 
For the sake of simplicity, we assume this relationship also holds here. 
SP-TCOs optical and electrical properties are thus controlled by only one parameter. 
Though there can be deviations from this relation, the general conclusions are not affected by this assumption (as shown in the Supplementary Fig.~9). 

In Fig.~\ref{fig:Transmittance_Conductivity}, next to the results for B-TCOs (purple and green), we plot the transmittance versus sheet conductivity for SP-TCOs with activation energies between 0.05 and 0.15~eV (in red, cyan and blue). 
Solid lines indicate carrier concentrations lower than $10^{22}$~cm$^{-3}$ while higher carrier concentrations are shown by dashed lines (as it is unlikely that higher carrier concentrations could be reached).
The transmittance dependence on the sheet conductivity for SP-TCOs is very different from the one for B-TCOs (purple and green). 
For SP-TCOs, the sheet conductivity can be enhanced by increasing the carrier concentration without degrading the transmittance up to concentrations as large as $10^{22}$~cm$^{-3}$. 
This directly originates from the very different physics of light absorption in SP- and B-TCOs, especially from the absence of plasma edge movement in the former.
 
\begin{figure}
	\center
	\includegraphics[clip,trim=0.2cm 0.3cm 0.3cm 0.1cm,width=.8\columnwidth]{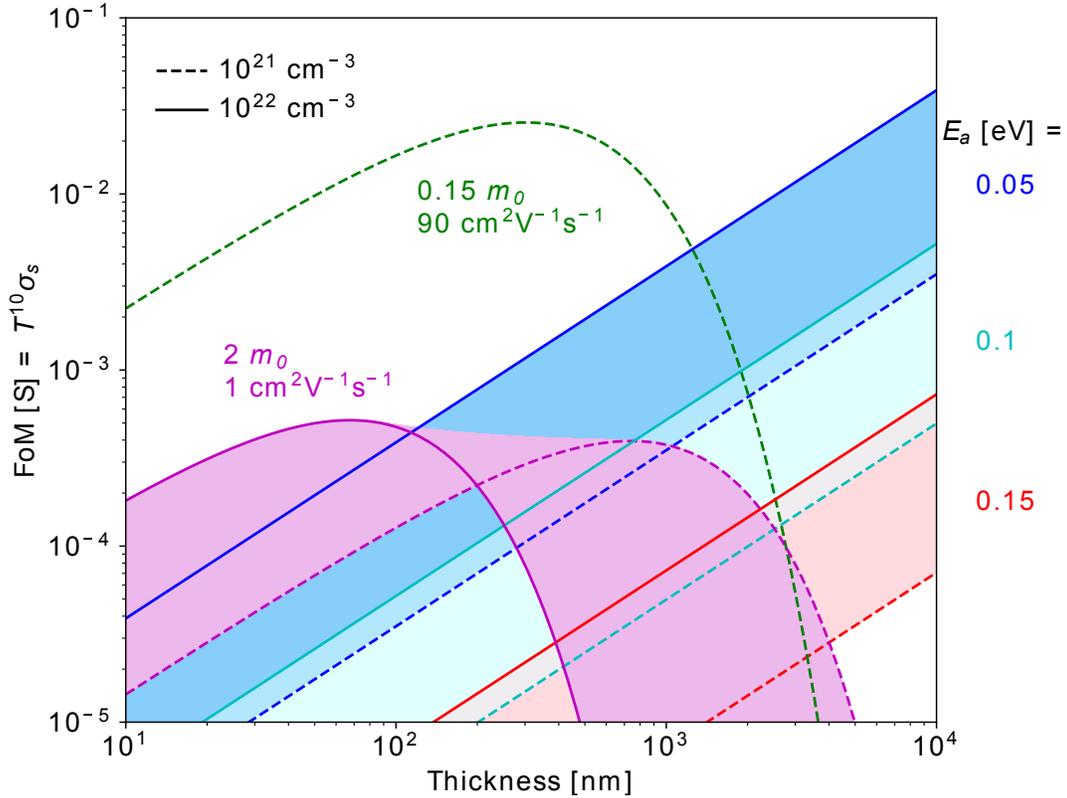}
	\caption{Haacke figure of merit (FoM) versus material thickness for typical n-type (green) and p-type (purple) B-TCOs, and SP-TCOs (blue, cyan and red) with various transport activation energies $E_a$. Dashed (resp. solid) lines correspond to a carrier concentration of $10^{21}$ (resp. $10^{22}$)~cm$^{-3}$. Concentrations between these two bounds are represented by colored zones.}
	\label{fig:FoM_thickness}
	\end{figure}

From Fig.~\ref{fig:Transmittance_Conductivity}, we can already observe that the SP-TCOs can offer attractive combined values of transparency and conductivity. 
However, the performances of TCOs are more easily compared using a figure of merit (FoM) which provides one number per material aggregating its optical and electrical responses~\cite{fleischer2017quantifying}.
Several FoMs exist and we chose the one suggested by Haacke as it is one of the most widely used~\cite{haacke1976new,gordon2000criteria,mendez2016figure,jacobs2016re}.
It is defined by $T^q \sigma_s$ with $T$ the transmittance, $\sigma_s$ the sheet conductivity and $q$ an arbitrary exponent usually set to 10 in the literature~\cite{ellmer2012past,zhang2016p,mendez2016figure,zhang2016correlated}.
Figure~\ref{fig:FoM_thickness} plots the FoM versus thickness for the low and high effective mass B-TCOs (green and purple, respectively) and SP-TCOs with different activation energies (blue, cyan and red). 
The dotted and solid lines indicate different doping levels ($10^{21}$ and $10^{22}$~cm$^{-3}$), and the shaded zones between them correspond to intermediate doping levels between these extremes.
For each B-TCO, an optimal thickness maximizes the FoM~\cite{ellmer2012past}.
The high and low effective mass B-TCOs show similar behavior of the FoM versus thickness with optimal thicknesses from 100~nm to \unit{1}{\micro\meter} and maximal FoMs respectively of $5\times 10^{-3}$ and $3\times 10^{-2}$ S. 
This explains the small thicknesses currently used in applications \cite{ellmer2012past}.
On the other hand, SP-TCOs (blue, cyan and red) do not show an optimal thickness. 
As a result they can reach a similar level of FoM ($3\times 10^{-2}$~S) as low effective mass/high mobility B-TCOs (in green), when they combine a very low activation energy (0.05~eV) with high carrier concentrations ($10^{22}$~cm$^{-3}$), and if they are thick enough (around \unit{10}{\micro\meter}). 
What is even more remarkable is that the SP-TCOs outperform by at least one order of magnitude in FoM the high effective mass/low mobility B-TCOs. At a thickness of \unit{5}{\micro\meter} (which is compatible with various applications such as solar cells~\cite{liu2010transparent,ma2011tco,bernal2016design,boccard2010transparent}), SP-TCOs with an activation energy of at most 0.1~eV and a doping between $10^{21}$ and $10^{22}$~cm$^{-3}$ will show larger FoM than low mobility B-TCOs of optimal thickness (100~nm -- \unit{1}{\micro\meter}). 
To put it simply, an SP-TCO with adequate materials properties (high doping level and low activation energy) will perform better than a lousy (low mobility/high effective mass) B-TCO. 
This conclusion is exacerbated in applications where transparency in the infra-red is key (e.g., solar cells)~\cite{liu2016effects,calnan2010high,yu2012ideal}.
Indeed, it is the plasma edge that is detrimental to IR transparency of highly doped B-TCOs and, in contrast, SP-TCOs with their absence of plasma edge offer excellent performances in the IR (see Supplementary Information, Section 4).

The outperformance of low effective mass B-TCOs by SP-TCOs is especially relevant for p-type TCOs. 
Indeed, n-type B-TCOs are currently available with very low effective masses and high mobilities. 
Materials such as In$_2$O$_3$, ZnO or SnO$_2$ are well represented by the low effective mass green curve in Figs.~\ref{fig:Transmittance_Conductivity} and \ref{fig:FoM_thickness}. 
In this case, there is no strong incentive to favor small-polaron transport. 
However, the situation for p-type TCOs is drastically different with no high mobility/low effective mass p-type B-TCO available. 
In fact, the oxygen $p$ character of the valence band in typical oxides is a major obstacle to the development of low effective mass p-type oxides. 
Data mining of large electronic structure databases have shown that it is extremely rare for a p-type transparent oxide to provide effective masses as low as current n-type TCOs~\cite{hautier2013identification,varley2017high}.
Keeping in mind this tendency for large effective masses in p-type oxides, our analysis justifies the search for new high performance p-type TCOs to turn to SP-TCOs. 

\FloatBarrier
\subsection*{New design route for p-type TCOs and the case of Sr-doped LaCrO$_3$}

We have outlined the general principle that SP-TCOs can outperform high effective mass/low mobility transparent oxides. 
However, as for B-TCOs, there are certain material properties that are required to lead to a high-performance SP-TCO. 
Our analysis can be directly used to outline a series of materials design principles, helping the discovery of novel efficient SP-TCOs. 
The first requirement, which is already present for B-TCOs, is a large band gap (ideally $>$ 3~eV). 
Next, the SP-TCO needs to present excitation energies not affecting the visible transparency. 
Typically, polaronic transition energies lower than 1.2~eV are required. 
It is also critical to reach decent mobilities with a reasonably low activation energy~\cite{exarhos2007cation}.
Our analysis indicates that polaron activation energies from 0.15~eV to 0.05~eV are required.
Finally, high dopability is even more important than in B-TCOs as the lower mobility needs to be compensated by higher carrier concentrations (typically $10^{21}$ to $10^{22}$~cm$^{-3}$). 
We should note that not only the dopant solubility should be high but that dopants should not introduce defects in the gap as they would lead to unwanted absorption. 
Our analysis using Haacke's FoM also stresses that such an ideal material would require to be deposited at thicknesses above the micrometer to maximize performances.
   
Cr-based oxides such as Cr$_2$O$_3$ and CuCrO$_2$ have been shown to involve small-polaron transport and decent TCO properties~\cite{farrell2016synthesis,arca2011magnesium}.
Among those, the material with the highest reported performances is Sr-doped LaCrO$_3$ and we will use our analysis to rationalize its performances~\cite{zhang2015perovskite,zhang2015hole}.
We perform this comparison through a combination of previously reported experimental data on Sr-doped LaCrO$_3$ as well as our own first-principles computations. 
Details are presented in Supplementary Information, Section 6.
We computed from first principles a transition energy ranging between 0.56 and 0.79~eV in very good agreement with the experimental absorption peak observed by Zhang \textit{et al.} at $0.7-0.8$~eV~\cite{zhang2015perovskite}.
Our computations assign this peak to the transition from the small-polaron state to the delocalized state. 
The computed activation energies for the hopping mechanism are $\sim$ 0.1~eV. 
The full data and the energy curves of the holes transfers are presented in Supplementary Information, Section 6.
The values are in good agreement with other theoretical work and experimental transport measurements~\cite{zhang2015hole}. 
Our computed energy of mixing for Sr in LaCrO$_3$ are close to zero indicating that entropy will easily favor the mixing of Sr in the oxide (see Supplementary Information, Section 6). 
Moreover, no defect states are introduced within the gap when Sr is incorporated in LaCrO$_3$.

LaCrO$_3$ appears to show all the design principles outlined previously and this rationalizes its good performances. 
However, the experimental Haacke's FoM of the best Sr-LaCrO$_3$ material lies several orders of magnitude lower than a perfect SP-TCO with the same activation energy (0.1~eV) and doping level ($3.4\times 10^{21}$~cm$^{-3}$)~\cite{zhang2015perovskite}.
The reason for this discrepancy comes from the band gap of LaCrO$_3$. 
Its experimental value is 2.8~eV~\cite{zhang2015perovskite,maiti1996electronic,sushko2013multiband} while our analysis assumes a material with a band gap higher than 3~eV.
LaCrO$_3$ will thus absorb in the end of the visible spectrum. 
Moreover, as Sr is inserted in the material, our computations show that the band gap shrinks and the transmittance drops (see Supplementary Information, Section 6) in agreement with the experimental observations~\cite{zhang2015perovskite}.
This analysis indicates that alternative materials to LaCrO$_3$ keeping the low activation energy for polaron hopping with slightly larger band gaps could lead to substantial improvement in terms of p-type TCOs performances.

Li-doped NiO is another example of p-type SP-TCO \cite{zhang2018electronic}. The increase of the hole concentration leads to an increase of the black color of NiO thin films. This effect can be due to different reasons. First, there might be a shrinkage of the band gap, leading to an increase of absorption coefficient, like in the case of LaCrO$_3$ and suggested by Ref.~\cite{chen2012nature}. Intrinsic defect states might also increase the absorption coefficient, as is suggested in Ref.~\cite{jang2010electrical}. Finally, there could be optical transitions between the small-polaron hole state and the O 2p6 - Ni 3d states in the valence band at an energy between 1.4 and 3 eV, as is suggested by Ref.~\cite{zhang2018electronic}. In any of these cases, Li-doped NiO does not fulfill the requirements to be an efficient small-polaron TCO.

Using well-known physical relations linking optical absorption and conductivity to material properties, we have shown that the transparency-conductivity compromise emerging from band carriers in B-TCOs does not exist when the transport is carried by small polarons. 
This important difference is linked to the absence of band carriers plasma absorption and reflection when the carriers are trapped and form small polarons. 
Our quantitative analysis of sheet conductivity versus transparency for TCOs exhibiting small-polaron or band-carrier transport shows that, in certain situations, SP-TCOs can outperform B-TCOs. 
This is especially the case when band carriers have large effective masses such as in typical p-type oxides. 
This result is surprising in view of the avoidance for small-polaron transport in the TCO field but rationalizes the recent emergence of relatively high-performance Cr-containing p-type TCOs (e.g., LaCrO$_3$). 
Our analysis leads to the outline of a series of materials design principles to develop high performance SP-TCOs. 
These design principles include a low activation energy for hopping ($<0.15$~eV) and high small-polaron concentrations (i.e., needing high dopant solubility) without defect states within the gap. 
We hope our analysis will motivate both experimental and computational searches for polaronic materials that could lead to very high performance TCOs especially p-type.

\section*{Methods}
\subsection*{Optical absorption and conductivity analysis}

For B- and SP-TCOs, we considered a single layer of TCO with perpendicularly incident solar light. 
In order to simplify the discussion, we did not take into account any interband transitions due to light absorption. 
The conclusions thus apply only for materials with a band gap $E_g > 3$~eV. 
We used the Drude model to derive the reflectivity $R$ and absorption coefficient $\alpha$ of B-TCOs as functions of the energy of the incoming light. 
We obtained the mean transmittance of visible light by integrating the transmittance of the solar black-body spectrum between 1.77 and 3.18~eV. 
The details and equations are reported in Supplementary Information, Section 1 and 2. 	

We used a simple Marcus model for the case of SP-TCOs~\cite{marcus1985electron,marcus1993electron}.
The system in a relaxed small-polaron state can absorb a photon to reach (a) a small-polaron excited state at a neighboring site or (b) the delocalized state, where the carrier is band-like~\cite{bjaalie2015small}.
More information on these transitions is given in Supplementary Information, Section 3.

\subsection*{Ab initio computations}
First principles density functional theory (DFT) calculations have been carried out with the Vienna Ab initio simulation package (VASP) as well as with the \textsc{abinit} software~\cite{gonze2016recent,amadon2008gamma}.
We used the generalized gradient approximation (GGA) of Perdew, Burke and Ernzerhof (PBE)~\cite{perdew1996generalized}.
A Hubbard correction $U$ (DFT+$U$) has been applied on the exchange-correlation functional to enable the localization of the hole~\cite{erhart2014efficacy}.
We chose $U = 3.7$~eV so that the computed band gap of LaCrO$_3$ corresponds to the experimental one of 2.8~eV~\cite{zhang2015perovskite,maiti1996electronic}.
All calculations have been performed with a kinetic energy cut-off of at least 520~eV for the plane-wave basis. 
We used a $4\times 4\times 4$ Monkhorst-Pack $k$-point grid for the Brillouin zone integrations in the case of the primitive cell and decreased the number of grid points when the cell size increases. 
The cell structural relaxation processes have been realized with a convergence parameter of 0.01~eV/$\angstrom$ for the maximum forces, and each self-consistent field step has been done with a convergence criterion of $10^{-7}$~eV on the total energy.
Polaron computations were performed using supercells and introducing holes (removing electrons). 
We checked for small-polaron formation by observing the magnetic moment on the Cr atoms as well as the Cr-O bond lengths.
The magnitude of the magnetic moment on the concerned Cr atom decreases by roughly 1 $\mu_B$ (Bohr magneton) when the hole is localized, effectively going from a Cr$^{3+}$ to a Cr$^{4+}$ state. 
We confirmed the antiferromagnetic configuration observed experimentally due to the antiparallel magnetic moments of the Cr atoms~\cite{sushko2013multiband,tseggai2008synthesis}.

The solid-state climbing image nudged elastic band (ss-cNEB) method has been used to compute the activation energy for small-polaron hopping~\cite{henkelman2000climbing,henkelman2000improved}.
We first interpolated the atomic positions between two neighboring small-polaron configurations of same spin orientation to obtain images along a migration path. 
The NEB method finds the minimum energy path between the first and the last images.

\bibliographystyle{naturemag}
\bibliography{References}
\nocite{momma2011vesta}
\nocite{ong2013python}

\section*{Acknowledgments}
G.B. and G.-M.R. acknowledge the F.R.S.-FNRS for financial support. Computational resources have been provided by the supercomputing facilities of the Universit\'e catholique de Louvain (CISM/UCL) and the Consortium des \'Equipements de Calcul Intensif en F\'ed\'eration Wallonie Bruxelles (C\'ECI) funded by the F.R.S.-FNRS. The present research benefited from computational resources made available on the Tier-1 supercomputer of the F\'ed\'eration Wallonie-Bruxelles, infrastructure funded by the Walloon Region under the grant agreement n\textsuperscript{$\circ$}1117545. 

\section*{Author contributions}
G.B. ran the first-principles computations and produced the figures. G.B. and G.H. wrote the article while all authors discussed and edited it.

\section*{Additional information}
\paragraph*{Supplementary Information} accompanies this paper.
\paragraph*{Competing financial interests:} The authors declare no competing financial interests.
\paragraph*{Reprints and permission}
\paragraph*{How to cite this article:}

\end{document}


\textbf{Are Small Polarons Always Detrimental for Transparent Conducting Oxides ?}

\begin{center}\begin{Large}{Supplementary Information}\end{Large}\end{center}

\textit{Guillaume Brunin, Gian-Marco Rignanese, and Geoffroy Hautier*}

\renewcommand{\figurename}{Supplementary Figure}
\renewcommand{\thefigure}{S\arabic{figure}}
\renewcommand{\theequation}{S\arabic{equation}}
\renewcommand{\tablename}{Supplementary Table}
\renewcommand{\thetable}{S\arabic{table}}

\section{Transmittance of thin films}
The transmittance $\tilde{T}(\omega)$ of a material of thickness $t$ is a function of the photon energy $\omega$ which is given by 
\begin{equation}
\tilde{T}(\omega) = (1-R(\omega))e^{-\alpha(\omega) t}.
\end{equation}
at normal incidence, where $R(\omega)$ and $\alpha(\omega)$ are its reflectivity and absorption coefficient. The former is given by
\begin{equation}
R(\omega) = \frac{(n(\omega)-1)^2 + \kappa(\omega)^2}{(n(\omega)+1)^2 + \kappa(\omega)^2} \label{eq:reflectivity}
\end{equation}
where $n(\omega)$ and $\kappa(\omega)$ are the refractive index and the extinction coefficient of the material. They are given by
\begin{equation}
n(\omega) = \sqrt{\frac{\varepsilon_1(\omega) + \sqrt{\varepsilon_1(\omega)^2+\varepsilon_2(\omega)^2}}{2}} \label{eq:refractive}
\end{equation}
\begin{equation}
\kappa(\omega) = \sqrt{\frac{-\varepsilon_1(\omega) + \sqrt{\varepsilon_1(\omega)^2+\varepsilon_2(\omega)^2}}{2}} \label{eq:extinction}
\end{equation}
where $\varepsilon_1(\omega)$ and $\varepsilon_2(\omega)$ are the real and imaginary parts of the dielectric function at the light frequency $\omega$. 
In order to compute the average transmittance of visible light, we use the spectral radiance $B_\lambda$ of a black body at a temperature of 5778 K to model the solar spectrum. The average transmittance is then computed as
\begin{equation}
T = \frac{\int_\mathrm{vis} B_\omega (1-R(\omega))e^{-\alpha(\omega) t} d\omega }{\int_\mathrm{vis} B_\omega d\omega}. \label{eq:transmittance}
\end{equation}

\section{Band-carriers TCOs : the Drude model}
\begin{figure}[h]
	\center
	\includegraphics[width=\columnwidth]{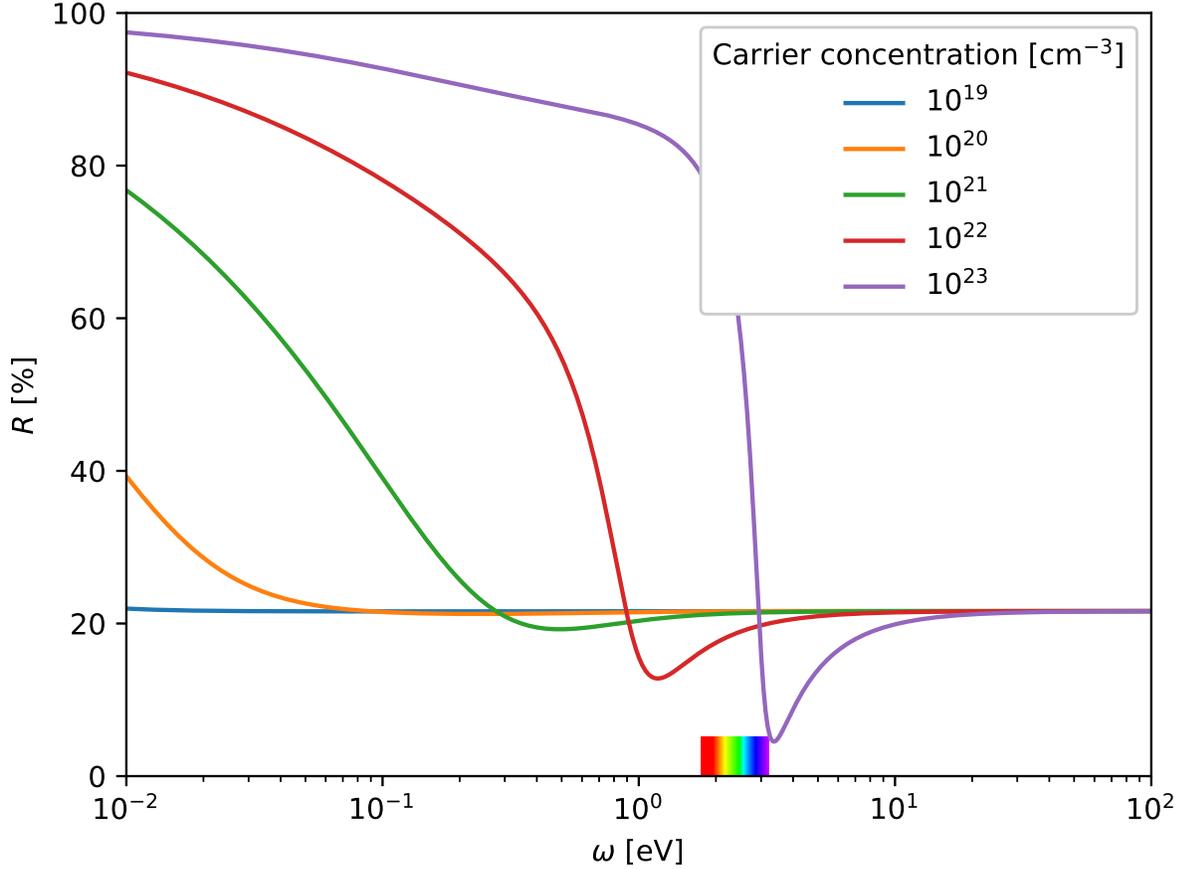}
	\caption{Reflectivity of a material based on the Drude model for different carrier concentrations. The mobility of carriers is $\mu = 1$~cm$^2$V$^{-1}$s$^{-1}$, their effective mass $m^* = 2 m_0$ and the high-frequency real part of the dielectric function $\varepsilon_r = 7.5$ . The visible spectrum ranges from 1.77 to 3.18~eV.}
	\label{fig:Reflectivity}
	\end{figure}

Charge carriers are free in most TCOs, meaning that they do not have a tendency to stay in a particular location of the material. The Drude model applies in this case and provides expressions for $\varepsilon_1(\omega)$ and $\varepsilon_2(\omega)$:~\cite{edwards2004basic,YuCardona}
\begin{equation}
\varepsilon_1(\omega) = \varepsilon_r - \frac{\omega_N^2 \tau^2}{1+\omega^2\tau^2}
\end{equation}
\begin{equation}
\varepsilon_2(\omega) = \frac{\omega_N^2 \tau}{\omega(1+\omega^2\tau^2)}
\end{equation}
with $\varepsilon_r$ the high-frequency limit of the real part of the dielectric function, $\omega_N^2 = Ne^2 / \varepsilon_0 m^*$, $N$ the carrier concentration, $e$ the elementary charge, $\varepsilon_0$ the vacuum permittivity, $m^*$ the carrier effective mass and $\tau$ their relaxation time. The plasma frequency $\omega_p$ is defined by
\begin{equation}
\omega_p^2 = \frac{\omega_N^2}{\varepsilon_r} = \frac{Ne^2}{\varepsilon_r \varepsilon_0 m^*}
\end{equation}
and gives the frequency under which light is reflected by the material, see Figure~\ref{fig:Reflectivity}. 

\begin{figure}[h]
	\center
	{\includegraphics[clip,trim=0.6cm 0.2cm 1.3cm 0.8cm,width=0.7\columnwidth]{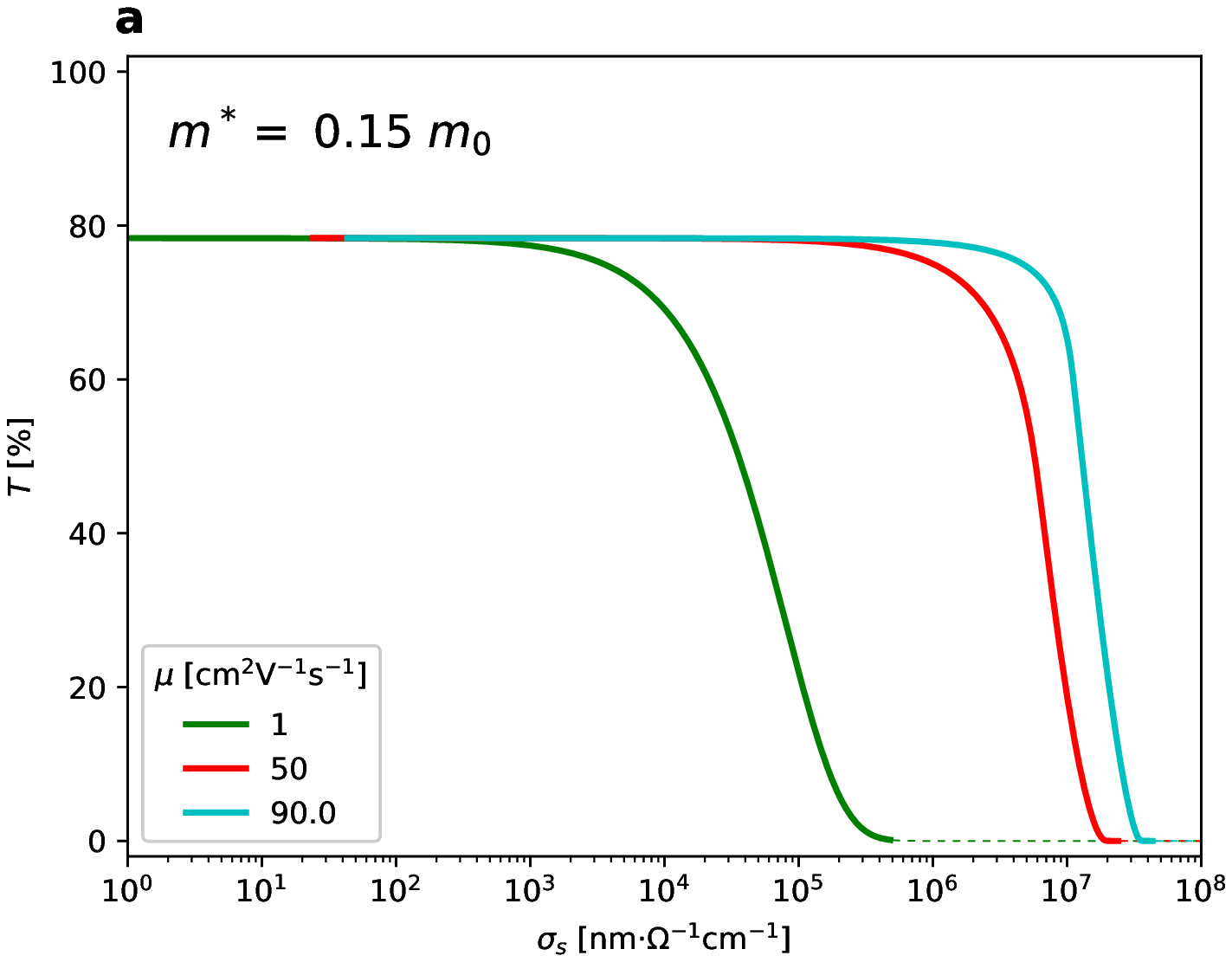}}
	
	\vspace{5mm}
	
	{\includegraphics[clip,trim=0.6cm 0.2cm 1.3cm 0.8cm,width=0.7\columnwidth]{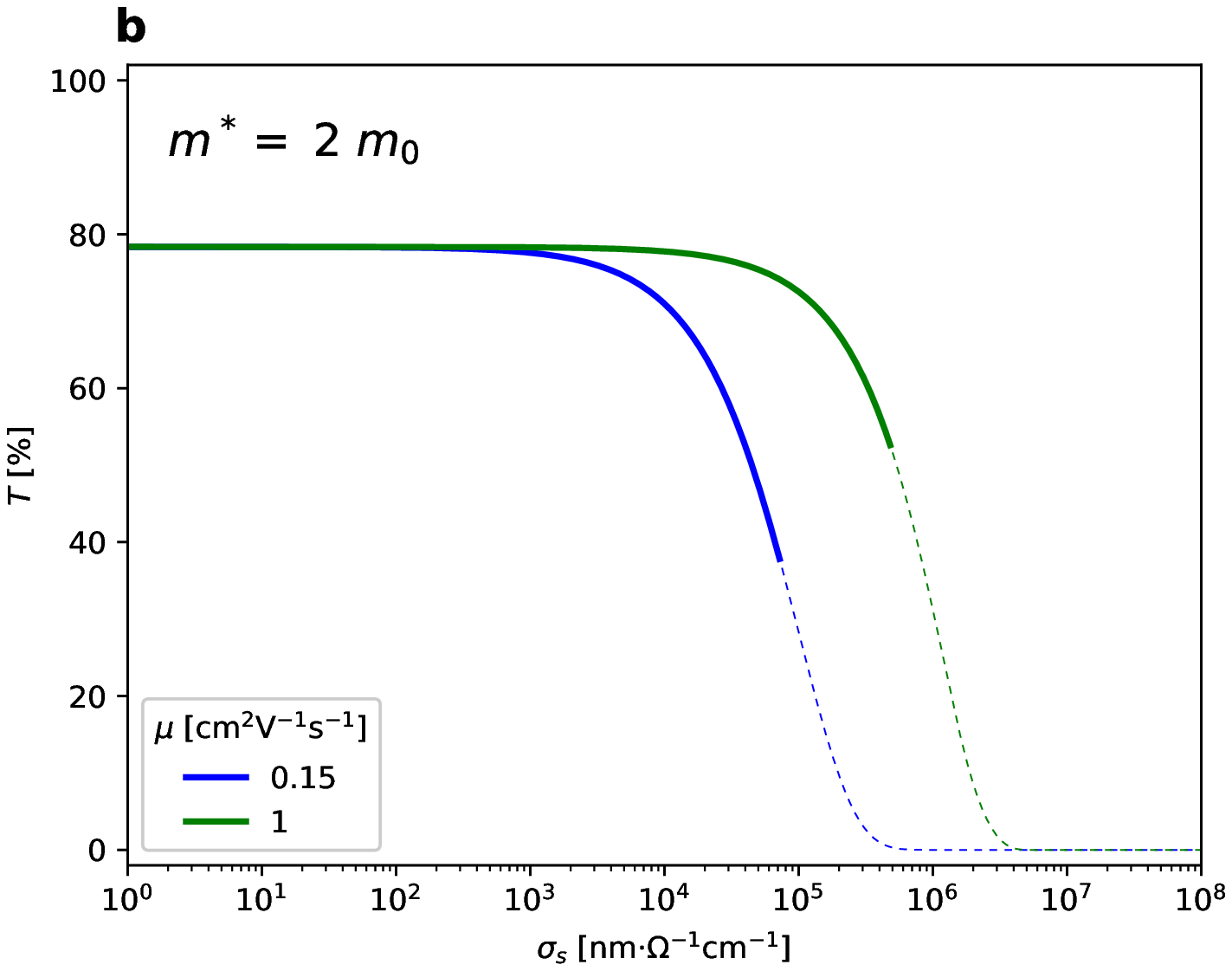}}
	\caption{Average transmittance of visible light through a B-TCO film of thickness $t = 300$~nm as a function of the sheet conductivity. Solid lines correspond to carrier concentration lower than $10^{22}$~cm$^{-3}$. Dashed lines correspond to a carrier concentration higher than $10^{22}$~cm$^{-3}$. The effective mass of the carriers is a) $0.15 m_0$ and b) $2 m_0$ and different mobilities are considered in each case.}
	\label{fig:Transmittance_Drude}
	\end{figure}

Because of the free-carrier absorption, B-TCOs usually have thicknesses around 100 nm - \unit{1}{\micro\meter} to keep a high transmittance \cite{ellmer2012past} (see Figure~3). 
The high-frequency real part of the dielectric function $\varepsilon_r$ does not play an important role in our discussion because its effect is the same on both types of materials: it defines the height of the high-reflectivity plateau. Moreover, the values of $\varepsilon_r$ found in the literature have the same order of magnitude.
We considered a thickness $t = 300$~nm, $\varepsilon_r = 7.5$ and the absorption coefficient of band carriers
\begin{equation}
\alpha(\omega) = \frac{2\omega\kappa(\omega)}{c} \label{eq:absorption_tot}
\end{equation}
to obtain Figure~\ref{fig:Transmittance_Drude}. The average transmittance is represented as a function of the sheet conductivity for two effective masses (typical of n-type in Figure~\ref{fig:Transmittance_Drude}a and of p-type in Figure~\ref{fig:Transmittance_Drude}b) and different mobilities (also typical values of n- and p-type in Figure~\ref{fig:Transmittance_Drude}a and \ref{fig:Transmittance_Drude}b respectively) in each case. As explained in the main text, when the sheet conductivity increases because of an increase of carrier concentration, the plasma edge shifts towards the visible part of the electromagnetic spectrum. At some point, depending on $m^*$ and $\mu$, the material becomes reflective and the transmittance drops.

\FloatBarrier
\section{Small-polaron materials}

\subsection{Absorption and reflectance} \label{supsec:decoupled}
\begin{figure}[h]
	\center
	{\includegraphics[clip,trim=0.4cm 0.3cm 1.5cm 0.6cm,width=0.7\columnwidth]{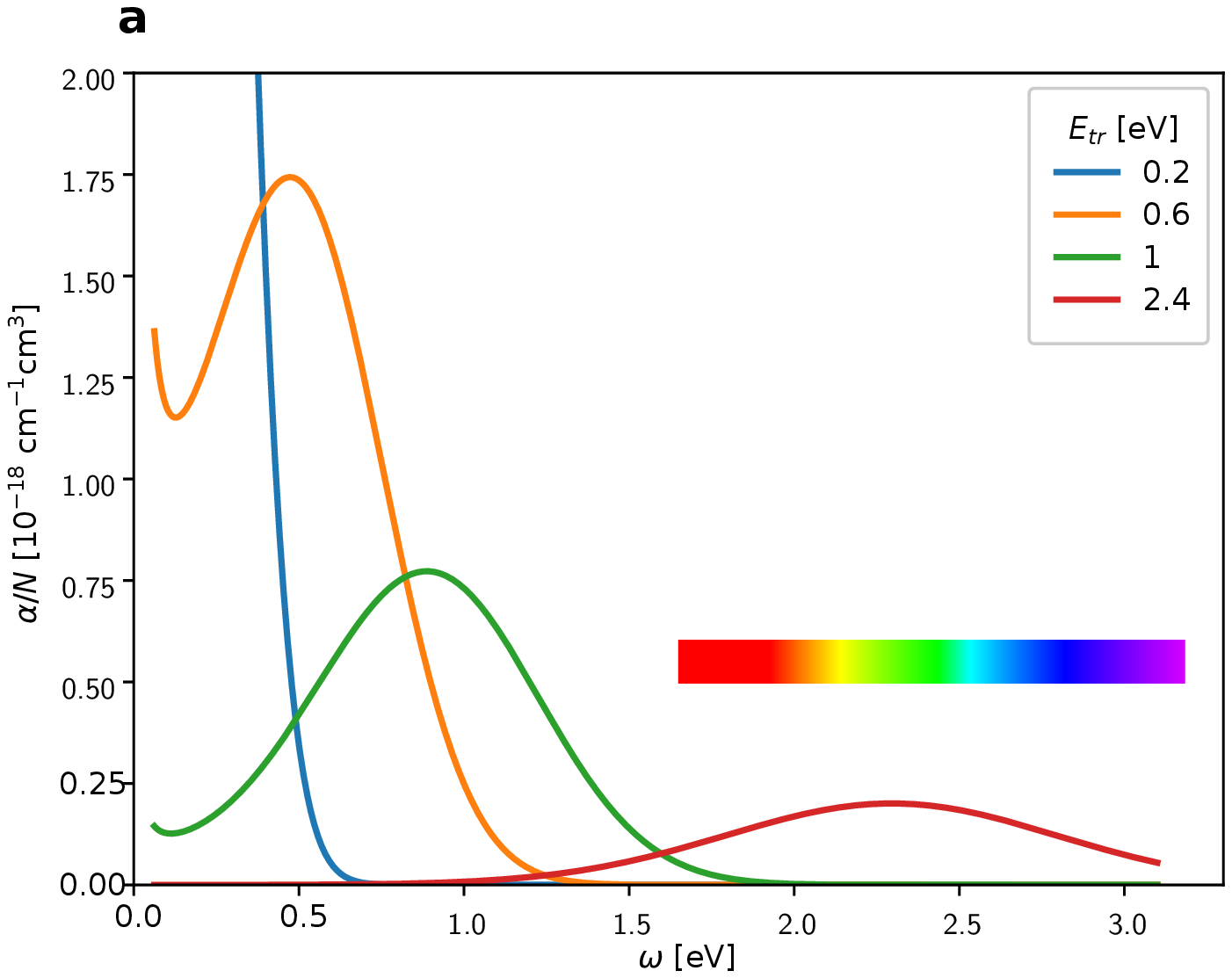}}
	
	\vspace{5mm}	
	{\includegraphics[clip,trim=0.4cm 0.2cm 1.5cm 0.6cm,width=0.7\columnwidth]{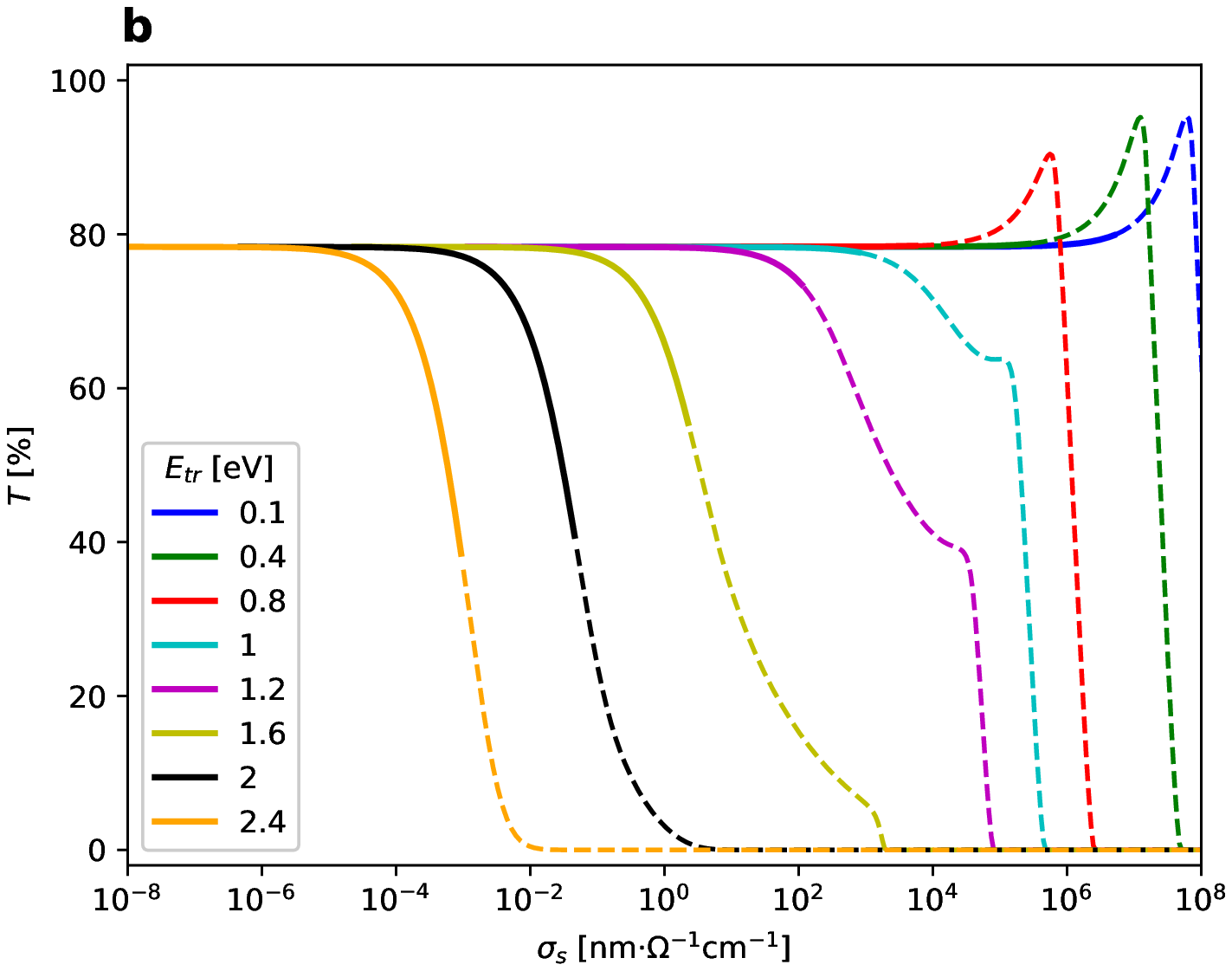}}
	\caption{a) Absorption coefficient of an SP-TCO for different transition energies. b) Average transmittance of visible light through a film of thickness $t = $~\unit{5}{\micro\meter} as a function of the sheet conductivity of an SP-TCO. Solid (resp. dashed) lines correspond to carrier concentrations lower (resp. larger) than $10^{22}$~cm$^{-3}$.}
	\label{fig:AT_Polaron}
	\end{figure}

	The small-polaron absorption coefficient defined in Equation~(2) is more precisely given by
	\begin{equation}
	\alpha(\omega) = N \frac{1}{\varepsilon_0} \frac{2\pi^{3/2}e^2}{m^\prime \omega c} \frac{J}{\Delta} \exp \left( \frac{-(E_{tr} - \hbar\omega)^2}{\Delta^2} \right)
	\end{equation}
	 with $N$ the small-polaron concentration, $\varepsilon_0$ the vacuum permittivity, $\omega$ the energy (or frequency) of the incoming light and $c$ its velocity, $m^\prime \equiv \hbar^2 / 2J a^2$, $J$ the intersite electronic transfer energy and $\Delta = \sqrt{8 E_p E_\text{vib}}$ the phonon broadening of the ground state with $E_p$ the small-polaron formation energy and $E_\text{vib}$ the vibrational energy (the largest value between $k_B T$ and $\hbar\omega_0/2$, the energy of the involved phonon mode), and $E_{tr}$ the optical transition energy. 
	We have set $\omega_0 = 600$~cm$^{-1}$ as a typical value, and therefore $E_\text{vib}$ = 40~meV.
	We used $J = 0.01$~eV in this work, a value computed for LaCrO$_3$. 
	This intersite energy is supposedly small in other materials as well. Moreover, its value (as well as the one of $m^\prime$) are not important as long as the transition energy is above 1.2~eV (see main text).
	The different transitions correspond to the excitation of the system to a neighboring small-polaron state (see Figure \ref{fig:Energy_Polaron}a) or to the delocalized state (Figure \ref{fig:Energy_Polaron}b).
	Figure~\ref{fig:AT_Polaron}a plots the absorption coefficient of an SP-TCO for different transition energies with typical parameters introduced in the text. The absorption peak is centered on the transition energy and its height decreases for increasing transition energies. 
	Using this absorption coefficient, we can obtain the extinction coefficient $\kappa(\omega)$ due to small polarons through Equation~\ref{eq:absorption_tot}. We then use the Kramers-Kronig relations to obtain the refractive index of the small-polaron system:
	\begin{equation}
	n(\omega) = n(\infty) + \frac{2}{\pi} P \int_{0}^\infty \frac{\omega^\prime \kappa(\omega^\prime)}{\omega^{\prime^2} - \omega^2} d\omega^\prime
	\end{equation}
	where $P$ indicates the principal value of the integral. As for B-TCOs, a lattice contribution needs to be added. In the high-frequency limit, we set $n(\infty) = \sqrt{\varepsilon_r}$ to reach the same limit as B-TCOs in the Drude model.

\begin{figure}[h]
	\center
	\includegraphics[clip,trim=0.7cm 0.2cm 0.2cm 0.1cm,width=0.75\columnwidth]{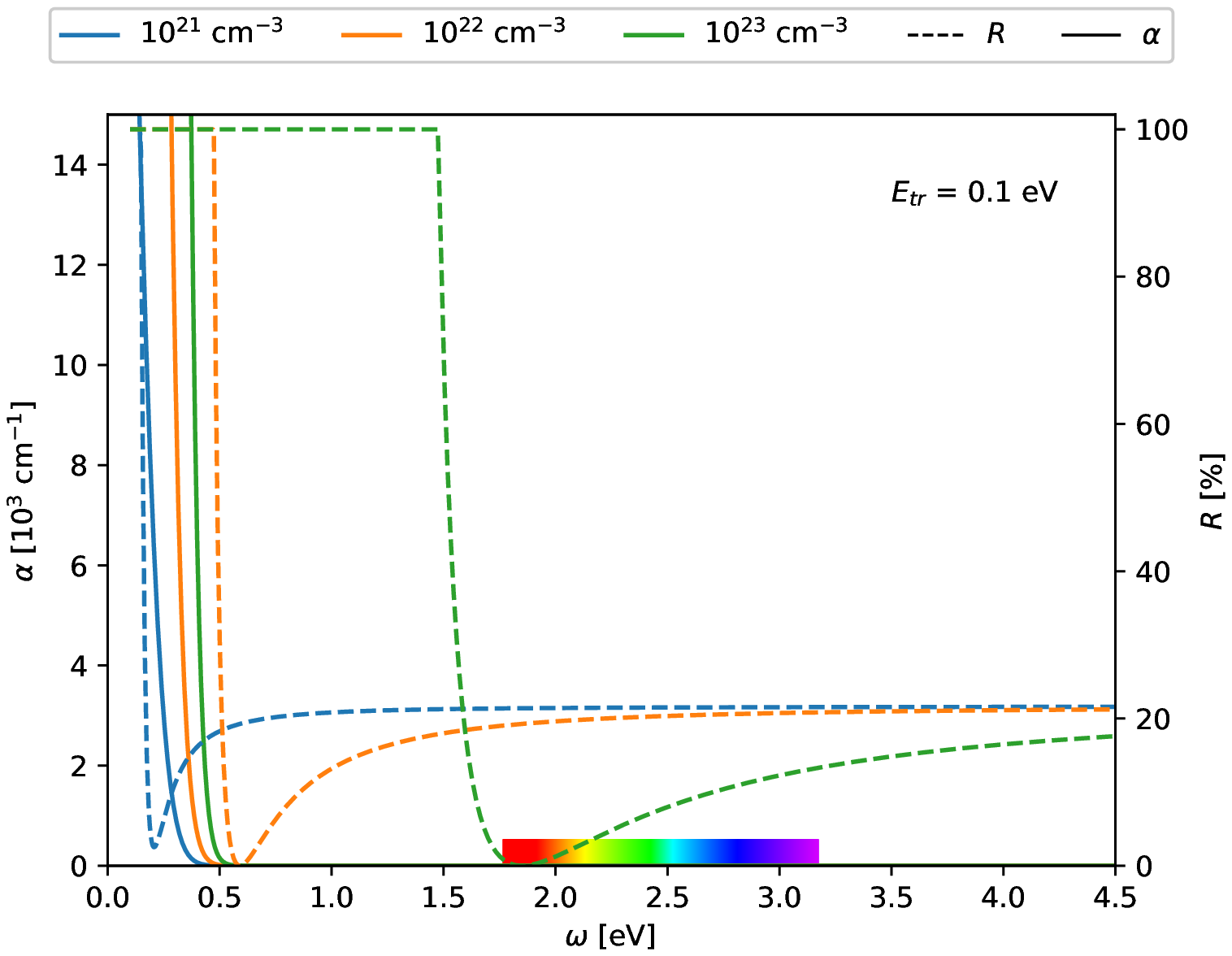}
	\caption{Absorption coefficient $\alpha$ and reflectivity $R$ of SP-TCOs with a transition energy of 0.1~eV. Three doping levels are considered : $10^{21}$, $10^{22}$ and $10^{23}$~cm$^{-3}$.}
	\label{fig:Optical_Pola0}
	\end{figure}

\begin{figure}[h]
	\center
	\includegraphics[clip,trim=0.7cm 0.2cm 0.2cm 0.1cm,width=0.75\columnwidth]{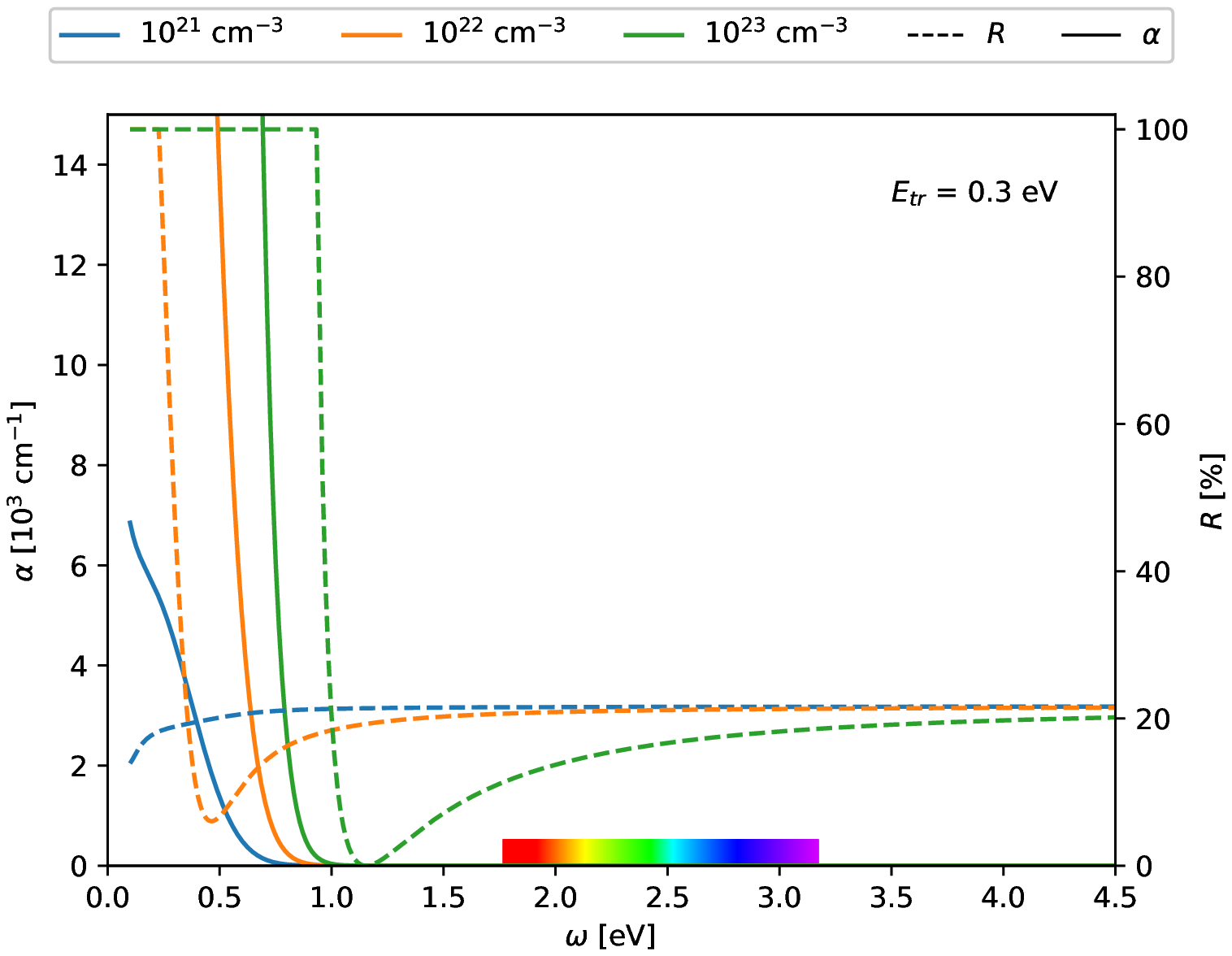}
	\caption{Absorption coefficient $\alpha$ and reflectivity $R$ of SP-TCOs with a transition energy of 0.3~eV. Three doping levels are considered : $10^{21}$, $10^{22}$ and $10^{23}$~cm$^{-3}$.}
	\label{fig:Optical_Pola2}
	\end{figure}

\begin{figure}[h]
	\center
	\includegraphics[clip,trim=0.7cm 0.2cm 0.2cm 0.1cm,width=0.75\columnwidth]{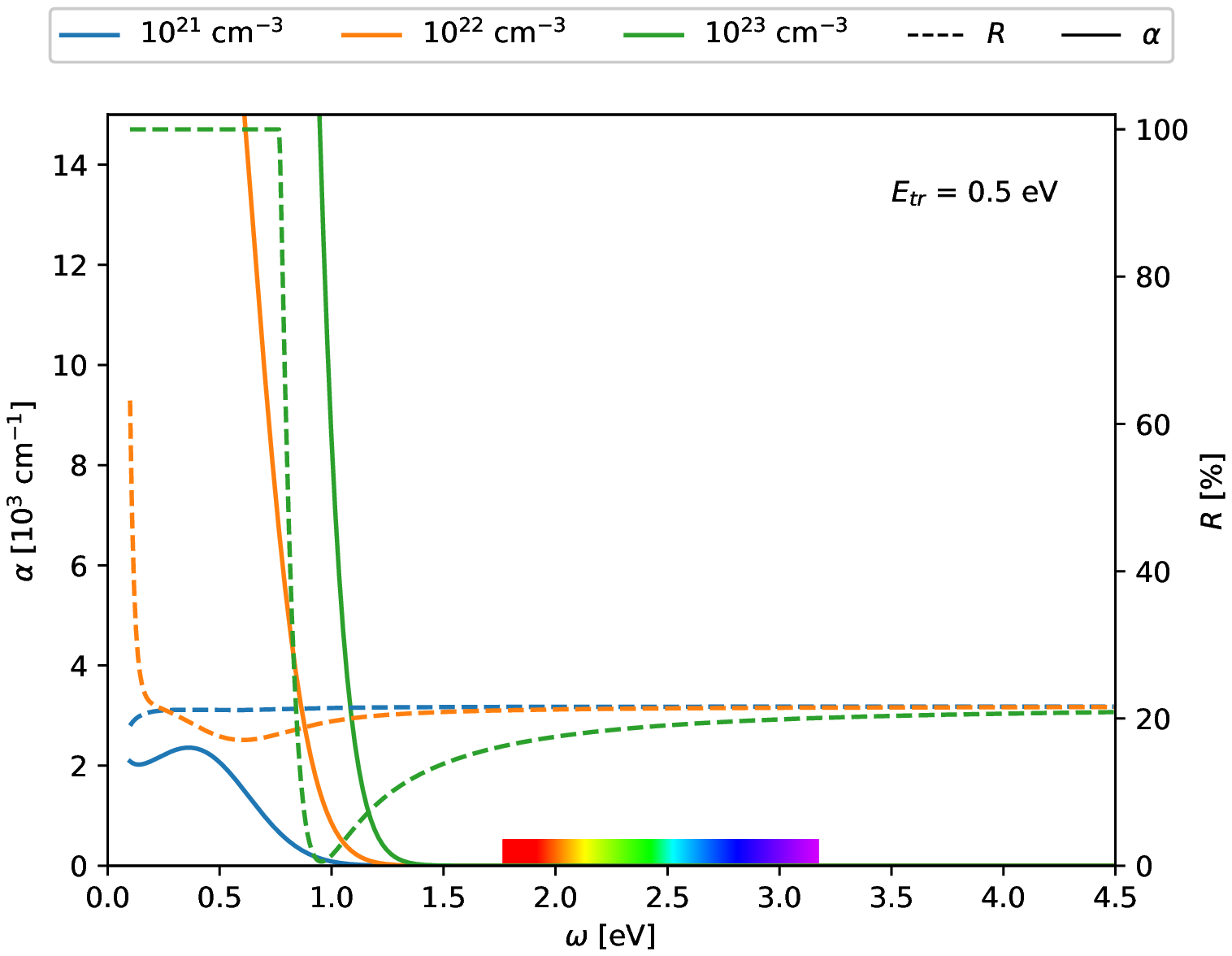}
	\caption{Absorption coefficient $\alpha$ and reflectivity $R$ of SP-TCOs with a transition energy of 0.5~eV. Three doping levels are considered : $10^{21}$, $10^{22}$ and $10^{23}$~cm$^{-3}$.}
	\label{fig:Optical_Pola4}
	\end{figure}

\begin{figure}[h]
	\center
	\includegraphics[clip,trim=0.7cm 0.2cm 0.2cm 0.1cm,width=0.75\columnwidth]{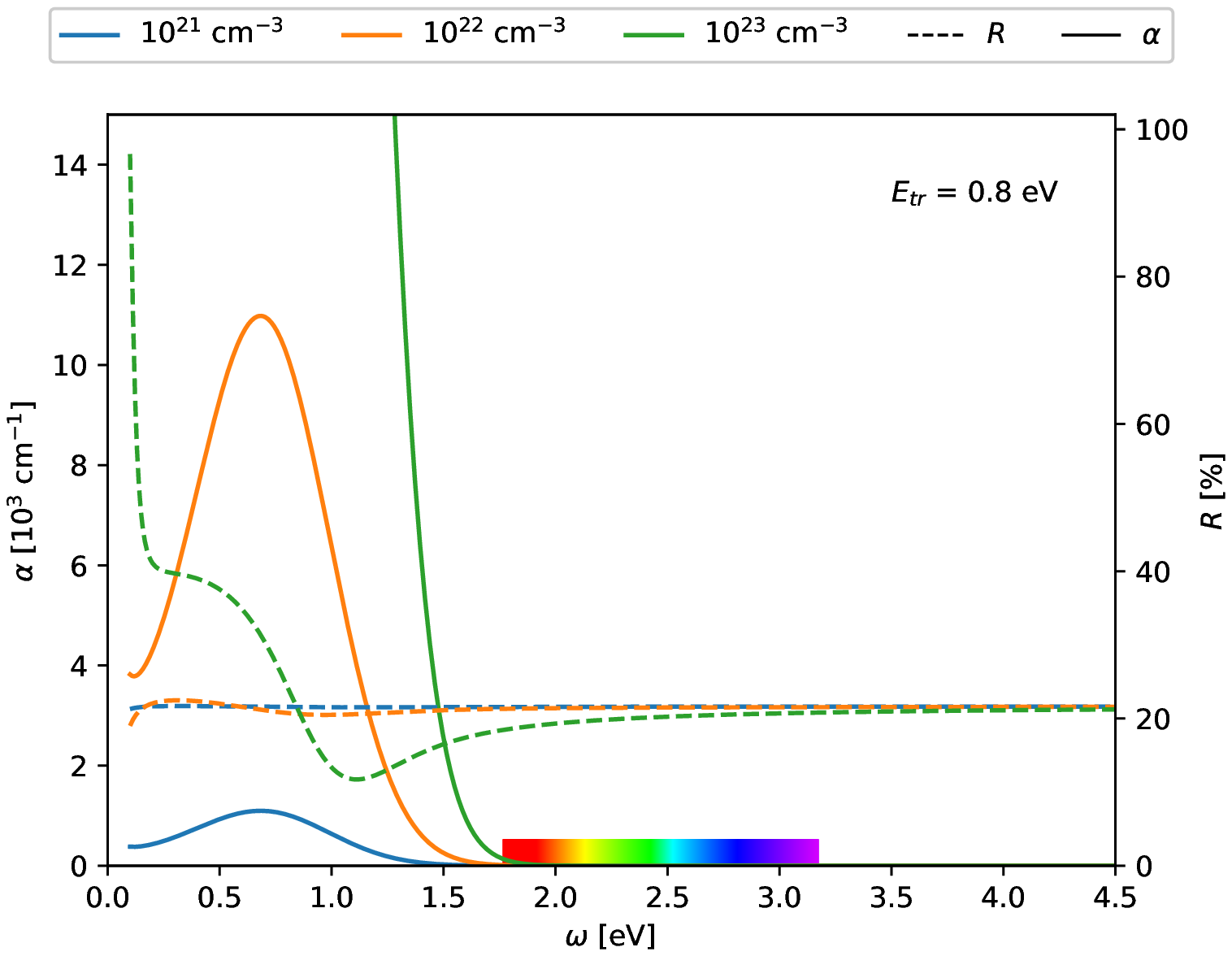}
	\caption{Absorption coefficient $\alpha$ and reflectivity $R$ of SP-TCOs with a transition energy of 0.8~eV. Three doping levels are considered : $10^{21}$, $10^{22}$ and $10^{23}$~cm$^{-3}$.}
	\label{fig:Optical_Pola7}
	\end{figure}

\begin{figure}[h]
	\center
	\includegraphics[clip,trim=0.7cm 0.2cm 0.2cm 0.1cm,width=0.75\columnwidth]{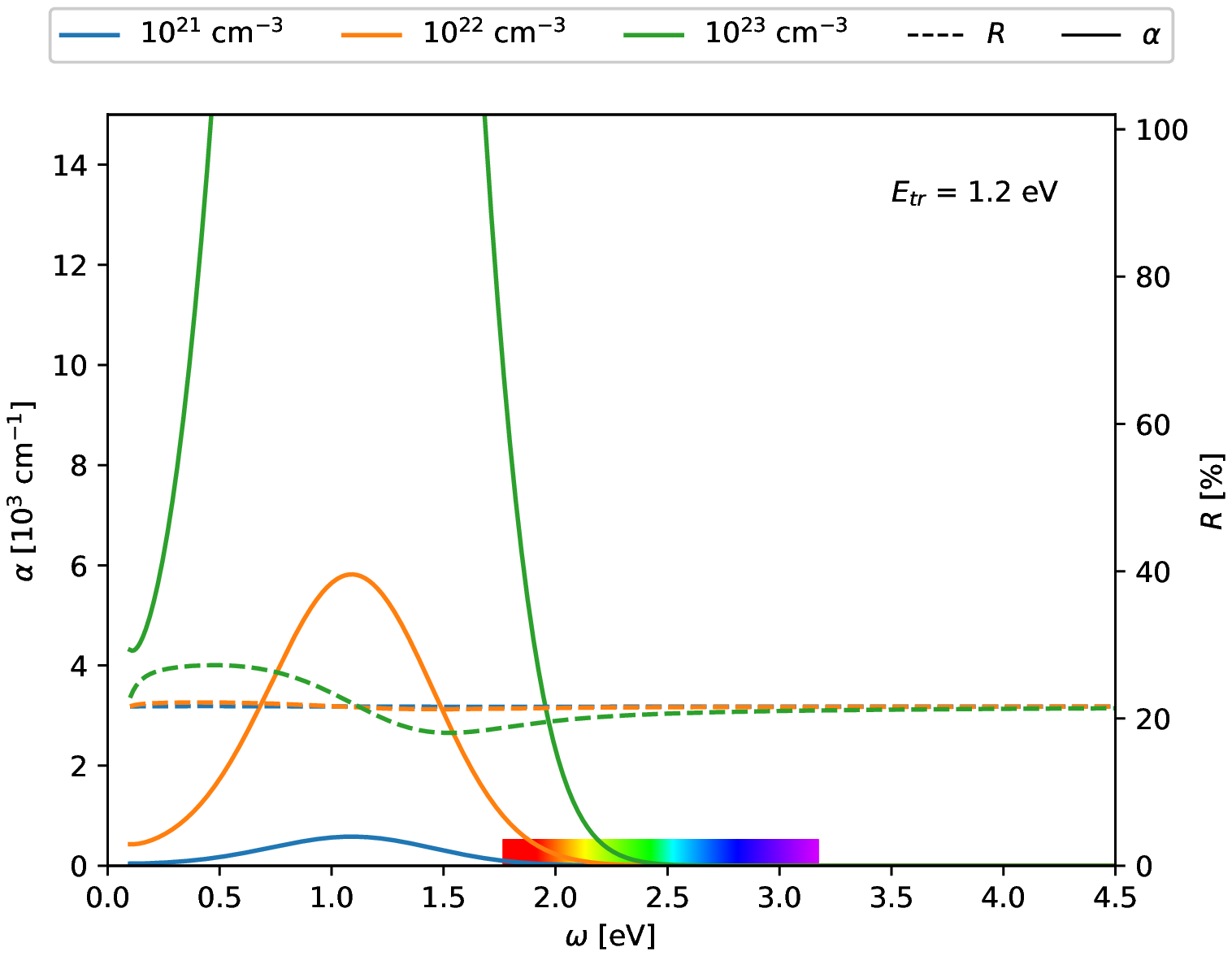}
	\caption{Absorption coefficient $\alpha$ and reflectivity $R$ of SP-TCOs with a transition energy of 1.2~eV. Three doping levels are considered : $10^{21}$, $10^{22}$ and $10^{23}$~cm$^{-3}$.}
	\label{fig:Optical_Pola11}
	\end{figure}

	Figure~\ref{fig:Optical_Pola0} to \ref{fig:Optical_Pola11} represent the total absorption coefficient together with the reflectivity of SP-TCOs for increasing transition energies and three doping levels.
	
\begin{figure}[h]
	\center
	\includegraphics[width=\columnwidth]{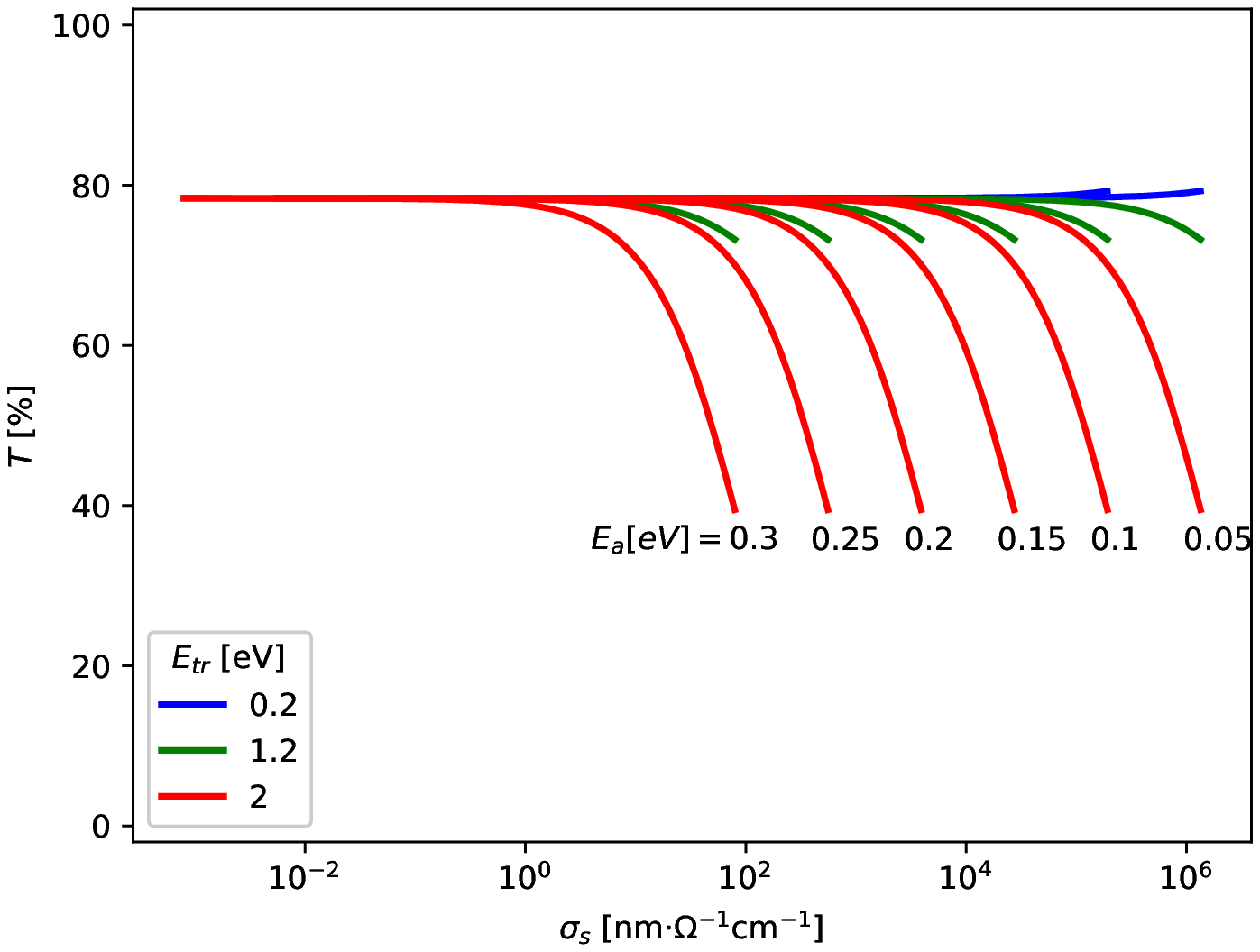}
	\caption{Average transmittance of visible light through a film of thickness $t = $~\unit{5}{\micro\meter} as a function of the sheet conductivity of an SP-TCO, with decoupled activation and transition energies. The transition energy influences only the transmittance, and the activation energy influences only the sheet conductivity.}
	\label{fig:Polaron_decoupled}
	\end{figure}

	The average transmittance $T$ of visible light has been obtained in the same way as for B-TCOs (Equation~\ref{eq:transmittance}) and is represented in Figure~\ref{fig:AT_Polaron}b versus sheet conductivity for different transition energies. The activation energy is in each case a quarter of the transition energy. A transition energy of 1.2~eV is necessary to observe a decrease of the transmittance for a carrier concentration under $10^{22}$~cm$^{-3}$. 
	We also considered the case where the transition energy is larger than four times the activation energy. The results are represented in Figure~\ref{fig:Polaron_decoupled}. As long as the transition energy is low enough, the activation energy is the most important parameter : the optical properties of SP-TCOs are not lessened for increasing carrier concentration, and the activation energy has a large impact on the mobility and hence the conductivity.

\subsection{Electrical properties}

\begin{figure}
	\center
	{\includegraphics[clip,trim=1.5cm 0.3cm 1.6cm 0.6cm,width=.5\columnwidth]{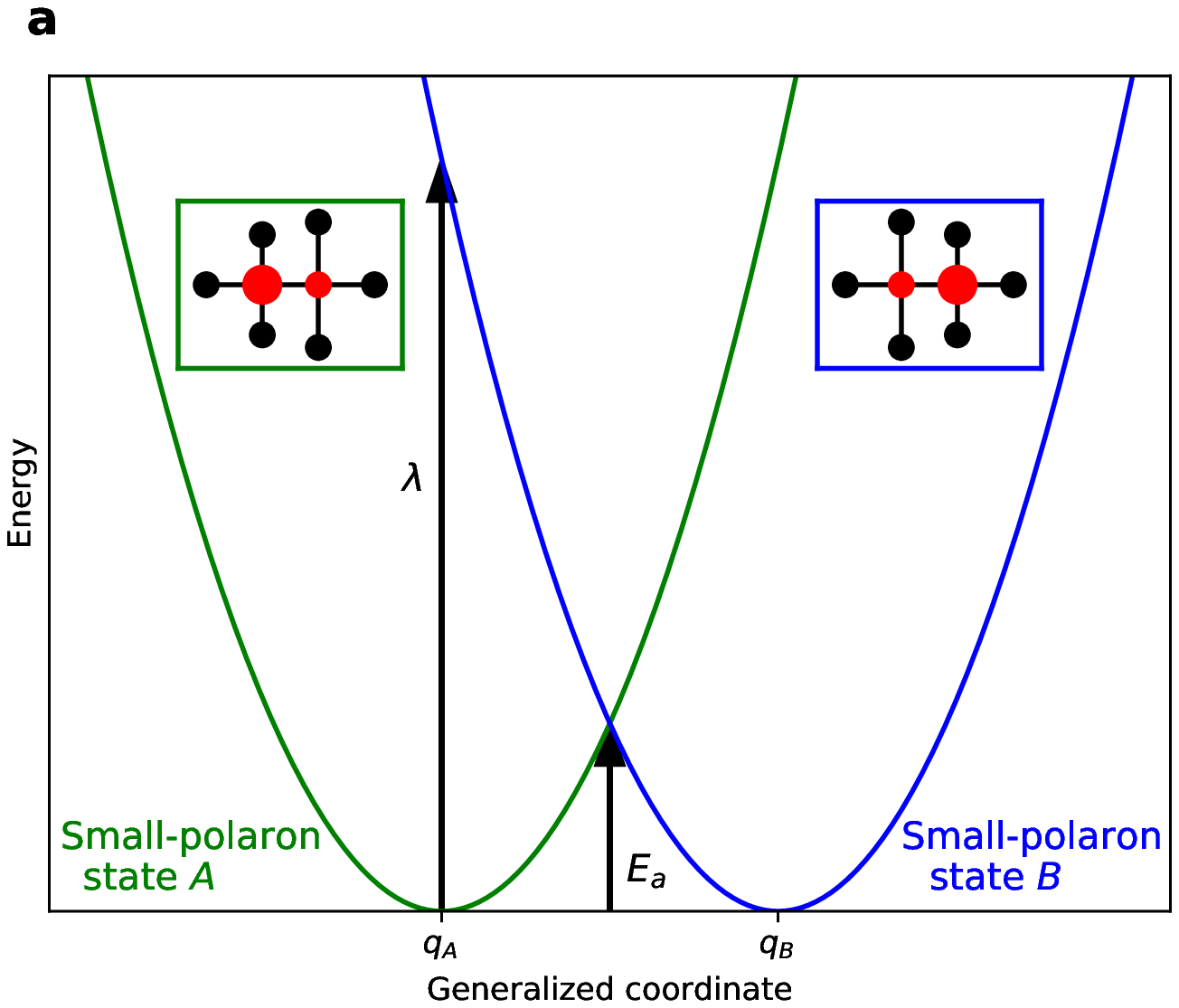}}
	
	\vspace{5mm}
	{\includegraphics[clip,trim=1.5cm 0.3cm 1.6cm 0.6cm,width=.5\columnwidth]{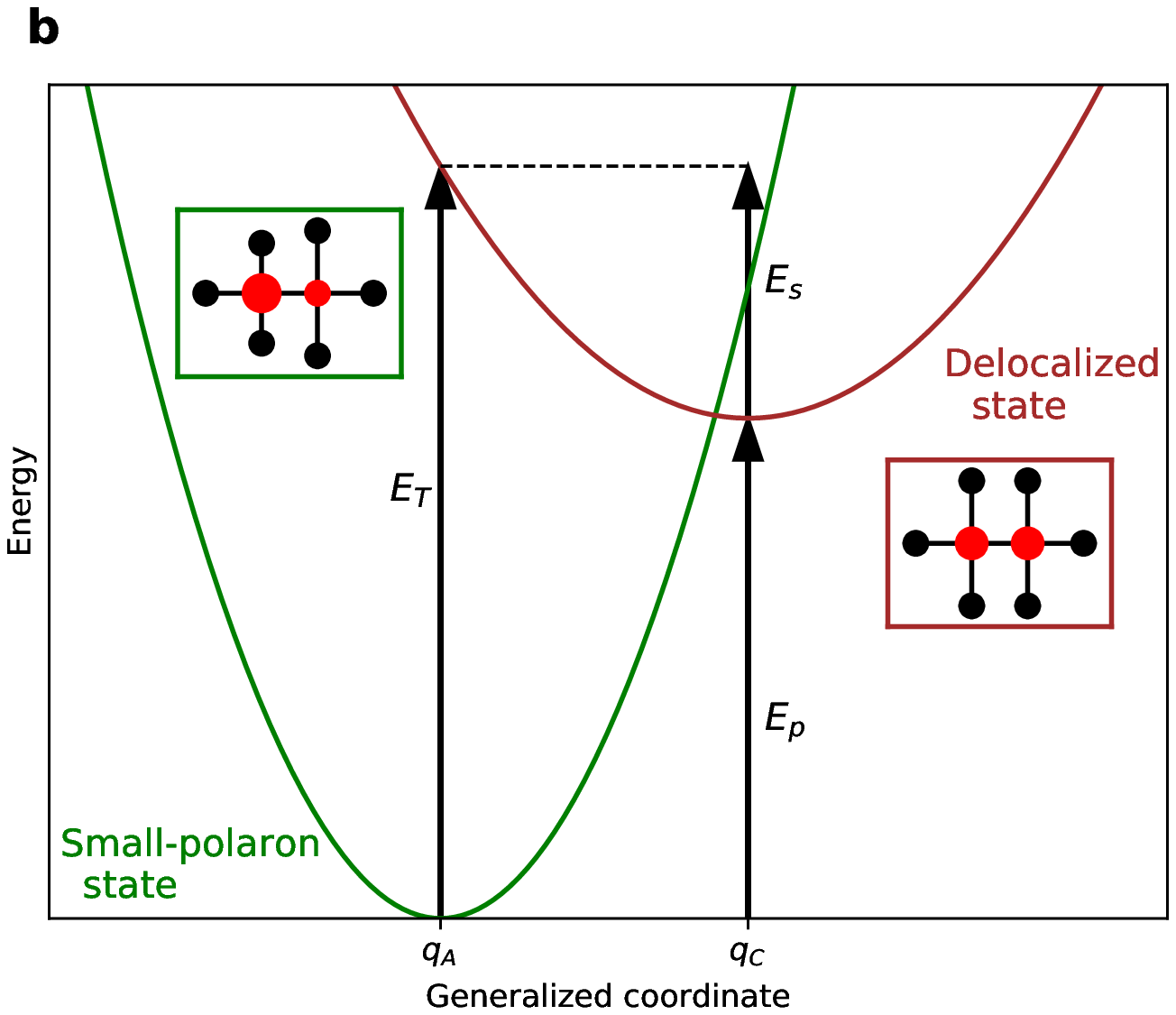}}
	\caption{Configuration-coordinate diagrams for a) the transport and the optical excitation of a small polaron to an adjacent site and b) the excitation of a small polaron to a delocalized state. $E_a$ is the activation energy, $\lambda$ the reorganization energy, $E_p$ the small-polaron formation energy, $E_s$ the strain energy, and $E_T$ the vertical transition energy. $q_A$ and $q_B$ are the generalized coordinates corresponding to the two neighboring small-polaron configurations. $q_C$ is the generalized coordinates corresponding to the delocalized configuration. The insets represent examples of structural configurations corresponding to $q_A$, $q_B$, $q_C$ in the simplest case of a two localizing atoms system. The charge localized on the red atoms is proportional to the size of the circles.}
	\label{fig:Energy_Polaron}
	\end{figure}

	The small-polaron hopping and the optical transitions of a small-polaron system are schematically represented in Figure~\ref{fig:Energy_Polaron}. The activation energy $E_a$, the small-polaron formation energy $E_p$, the reorganization energy $\lambda$ as well as the strain energy $E_s$ are defined.\cite{bjaalie2015small,himmetoglu2014interband}
	The activated mobility given by Equation (3) is explicitly written as
	\begin{equation}
	\mu = \frac{ea^2\omega_{LO}}{6k_BT} \exp\left(\frac{-E_a}{k_BT}\right) \label{eq:mobility_full}
	\end{equation}
with $e$ the elementary charge, $a$ the hopping distance, $\omega_{LO}$ the LO-phonon frequency, $k_B$ the Boltzmann constant and $T$ the temperature. We assumed room-temperature, and used $a = 4$~$\angstrom$ and $\omega_{LO} = 600$~cm$^{-1}$ as typical parameters. 
The hopping distance $a$ is the distance between two localization sites, and its value should be on the order of a few $\angstrom$. The value of $\omega_{LO}$ of different materials also usually have the same order of magnitude. The prefactor of the exponential in Equation~\eqref{eq:mobility_full} will not vary of orders of magnitude between different materials.

\subsection{Comparison between SP-TCOs and B-TCOs}
\begin{figure}[h]
	\center
	\includegraphics[clip,trim=0.6cm 0.5cm 1.6cm 1.9cm,width=0.7\textwidth]{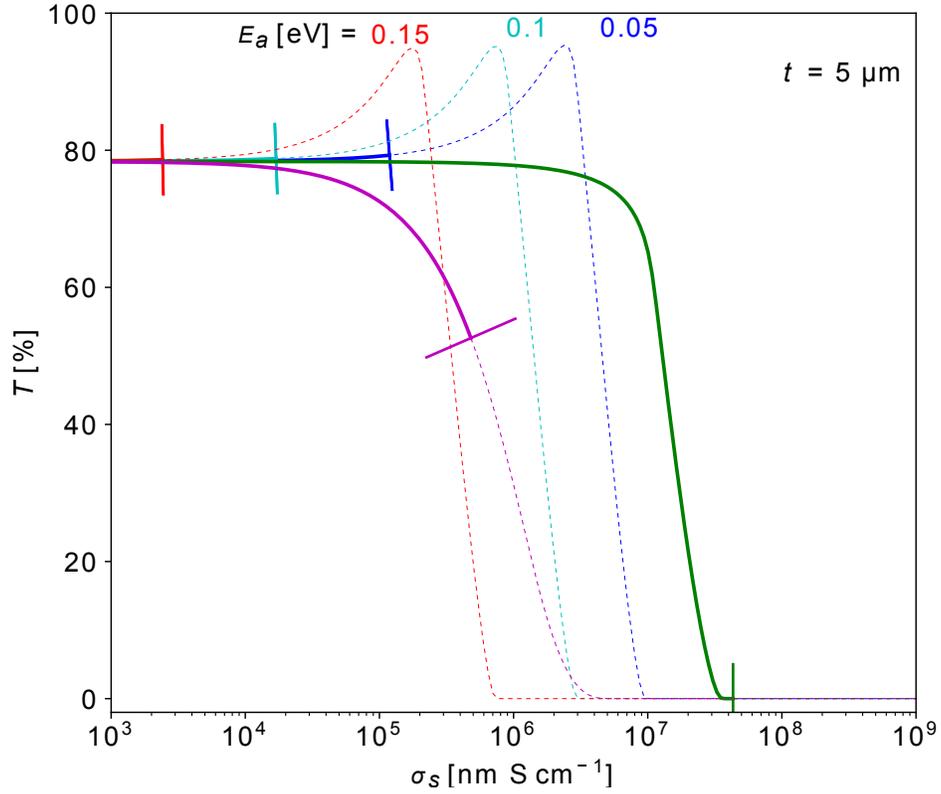}
	\caption{Average transmittance $T$ of IR and visible light versus sheet conductivity $\sigma_s$ for low effective mass/high mobility (green) and high effective mass/low mobility (purple) B-TCOs, and SP-TCOs (blue, cyan and red) with various transport activation energies $E_a$. We used a thickness $t = 300$ nm. Solid (resp. dashed) lines correspond to carrier concentrations lower (resp. higher) than $10^{22}$~cm$^{-3}$.}
	\label{fig:Transmittance_Conductivity_300nm}
	\end{figure}

In the main text, the comparison of the transmittance-conductivity curves of SP-TCOs and B-TCOs is done for a fixed thickness of \unit{5}{\micro\meter}. We can do the same analysis for a thickness of 300~nm. Fig.~\ref{fig:Transmittance_Conductivity_300nm} shows the comparison between the two types of materials. 

\FloatBarrier
\section{Comparison between band and small-polaron materials including the infrared}
	
\begin{figure}[h]
	\center
	\includegraphics[clip,trim=0.6cm 0.5cm 1.6cm 1.9cm,width=0.7\textwidth]{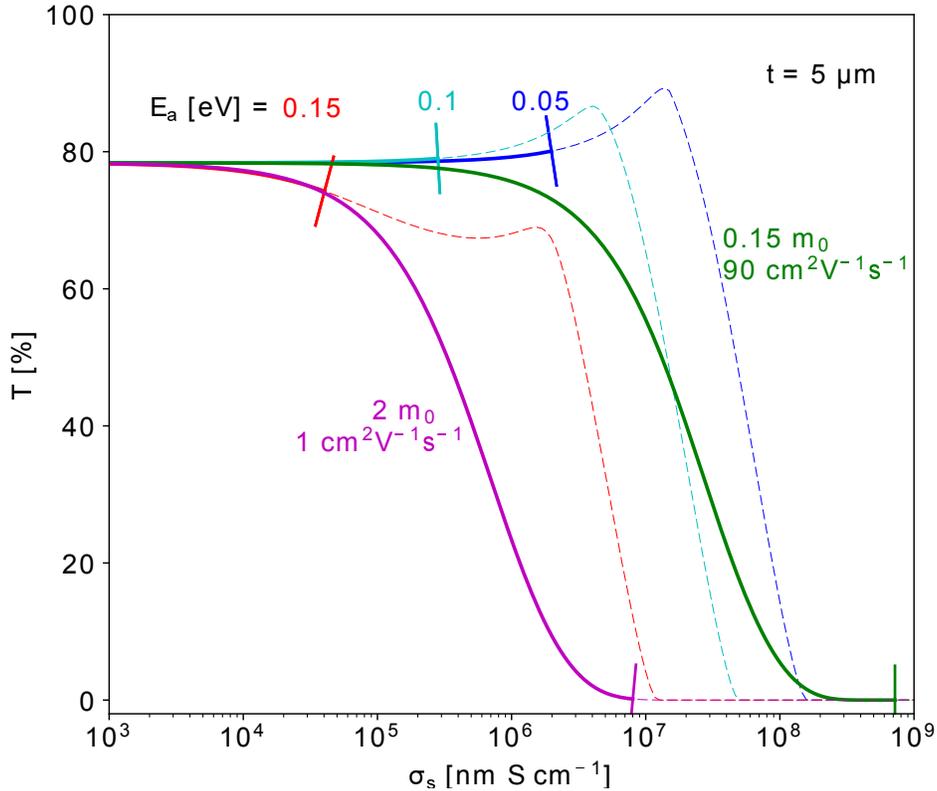}
	\caption{Average transmittance $T$ of IR and visible light versus sheet conductivity $\sigma_s$ for low effective mass/high mobility (green) and high effective mass/low mobility (purple) B-TCOs, and SP-TCOs (blue, cyan and red) with various transport activation energies $E_a$. We used a thickness $t = $~\unit{5}{\micro\meter}. Solid (resp. dashed) lines correspond to carrier concentrations lower (resp. higher) than $10^{22}$~cm$^{-3}$.}
	\label{fig:Transmittance_Conductivity_IR+VIS}
	\end{figure}

\begin{figure}[h]
	\center
	\includegraphics[clip,trim=0.6cm 0.4cm 1.6cm 1.9cm,width=0.7\textwidth]{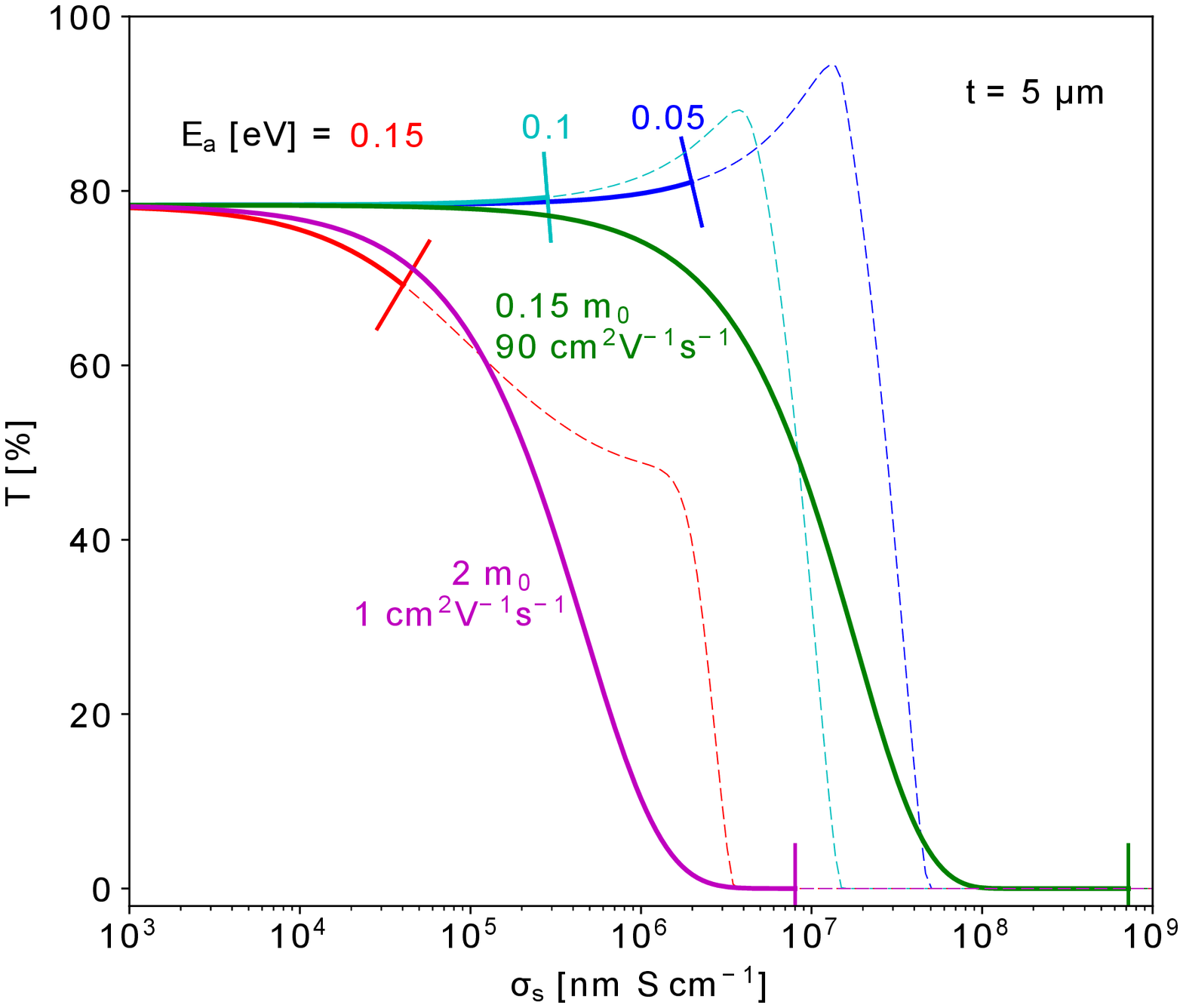}
	\caption{Average transmittance $T$ of IR versus sheet conductivity $\sigma_s$ for low effective mass/high mobility (green) and high effective mass/low mobility (purple) B-TCOs, and SP-TCOs (blue, cyan and red) with various transport activation energies $E_a$. We used a thickness $t = $~\unit{5}{\micro\meter}. Solid (resp. dashed) lines correspond to carrier concentrations lower (resp. higher) than $10^{22}$~cm$^{-3}$.}
	\label{fig:Transmittance_Conductivity_IR}
	\end{figure}

Another pertinent comparison between the two types of materials is to include the infrared (IR) or to consider this part of the electromagnetic spectrum only. We defined the IR as the light of energy comprised between 1 and 1.77~eV (1240 to 700~nm). Indeed, as mentioned in the main text, in solar cells the IR is also important and should be transmitted by the TCO layer. The problem with the B-TCOs is that the plasma edge can be located in the IR, and the TCO is then reflective for this part of the spectrum. SP-TCOs do not have this problem and are even more efficient in the IR if the optical transition energy is low enough. Figure~\ref{fig:Transmittance_Conductivity_IR+VIS} plots the average transmittance of visible light for the IR and the visible light combined, and Figure~\ref{fig:Transmittance_Conductivity_IR} plots the transmittance of IR only. We considered two thicknesses, one activation energy and typical n- and p-type B-TCOs. 

\begin{figure}[h]
	\center
	\includegraphics[clip,trim=0.2cm 0.3cm 0.3cm 0.2cm,width=0.8\textwidth]{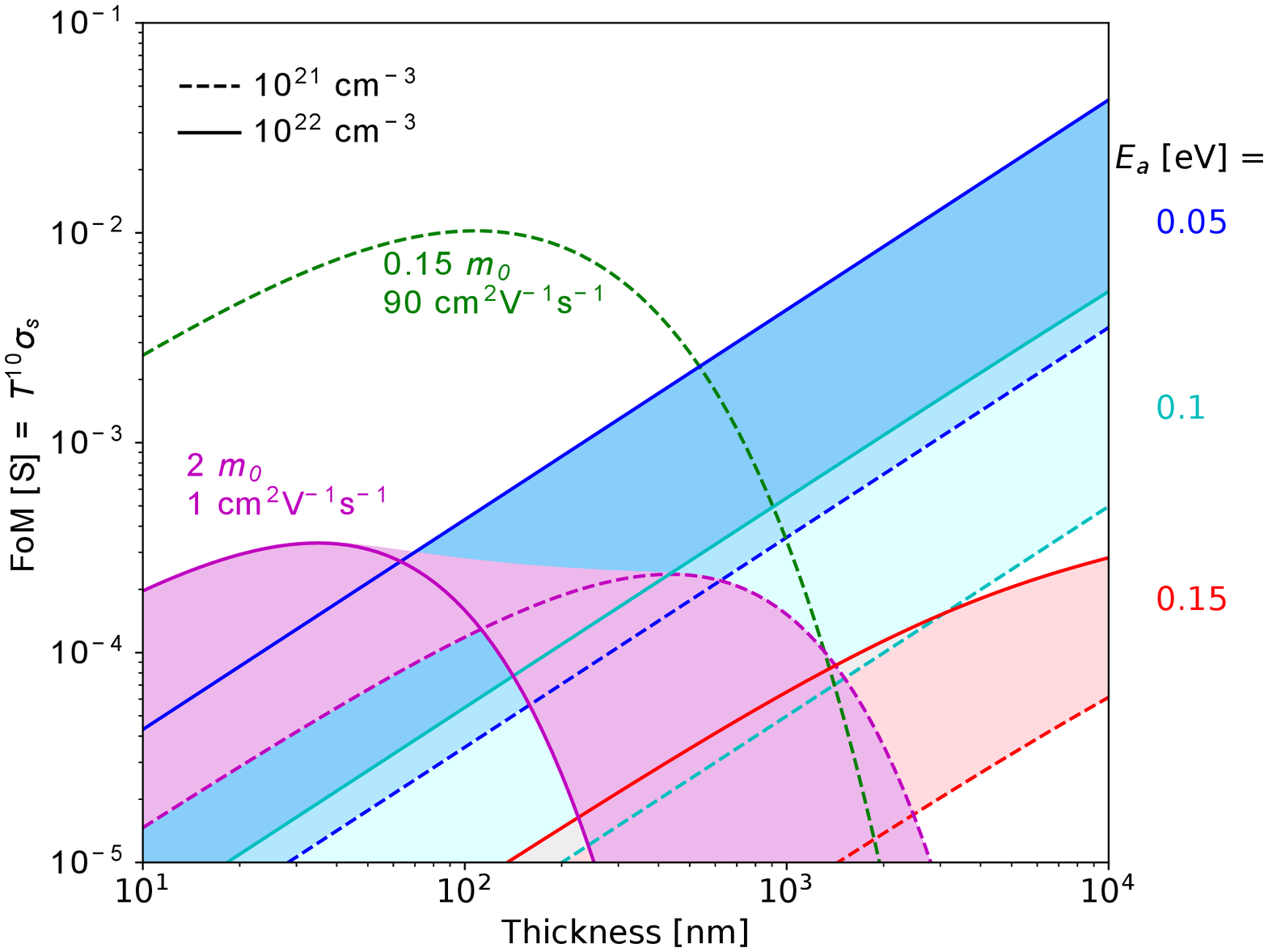}
	\caption{IR-including Haacke figure of merit (FoM) versus material thickness for typical n-type (green) and p-type (purple) B-TCOs, and SP-TCOs (blue, cyan and red) with various transport activation energies $E_a$. Dashed (resp. solid) lines correspond to a carrier concentration of $10^{21}$ (resp. $10^{22}$)~cm$^{-3}$. Concentrations between these two bounds are represented by colored zones.}
	\label{fig:FoM_Haacke_IR+VIS}
	\end{figure}

\begin{figure}[h]
	\center
	\includegraphics[clip,trim=0.2cm 0.3cm 0.3cm 0.2cm,width=0.8\textwidth]{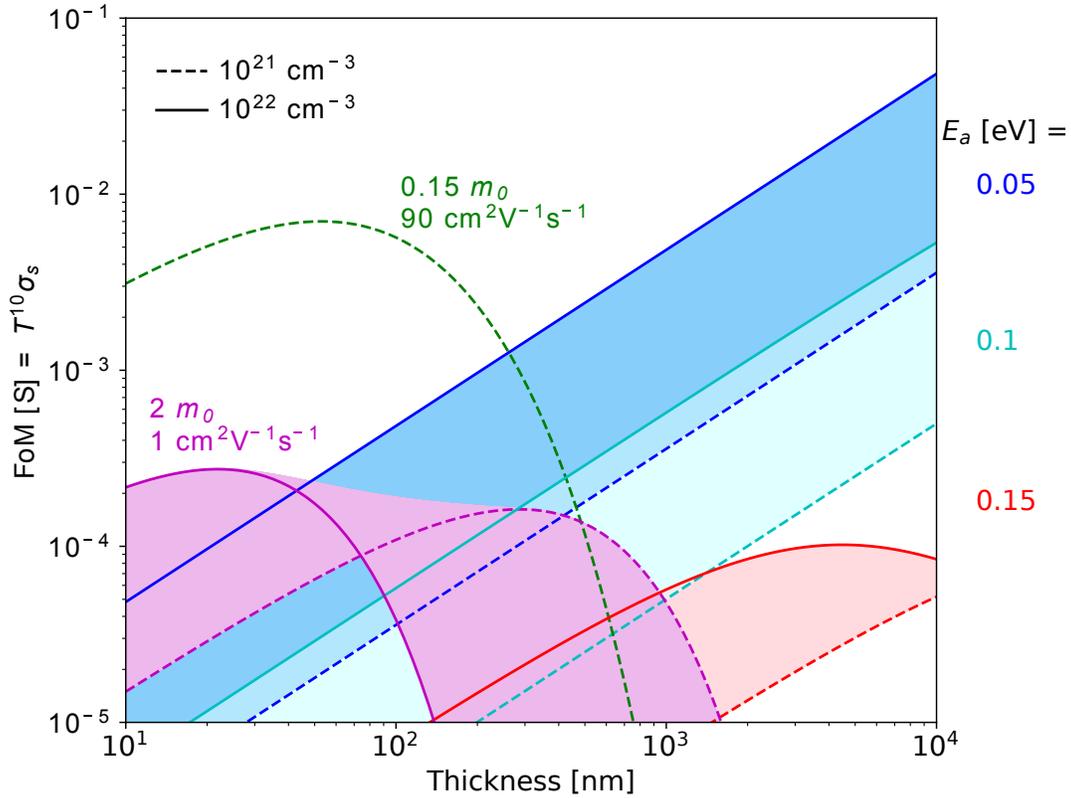}
	\caption{IR-only Haacke figure of merit (FoM) versus material thickness for typical n-type (green) and p-type (purple) B-TCOs, and SP-TCOs (blue, cyan and red) with various transport activation energies $E_a$. Dashed (resp. solid) lines correspond to a carrier concentration of $10^{21}$ (resp. $10^{22}$)~cm$^{-3}$. Concentrations between these two bounds are represented by colored zones.}
	\label{fig:FoM_Haacke_IR}
	\end{figure}

We can also compare the figure of merit including the IR (Figure~\ref{fig:FoM_Haacke_IR+VIS}) or considering only the IR (Figure~\ref{fig:FoM_Haacke_IR}). 
In the IR, the advantage of SP-TCOs over low mobility B-TCOs is exacerbated.
	
\FloatBarrier
\section{Comparison between different definitions of the figure of merit}

\begin{figure}[h]
	\center
	\includegraphics[clip,trim=2.1cm 2.8cm 2.4cm 3.5cm,width=0.75\textwidth]{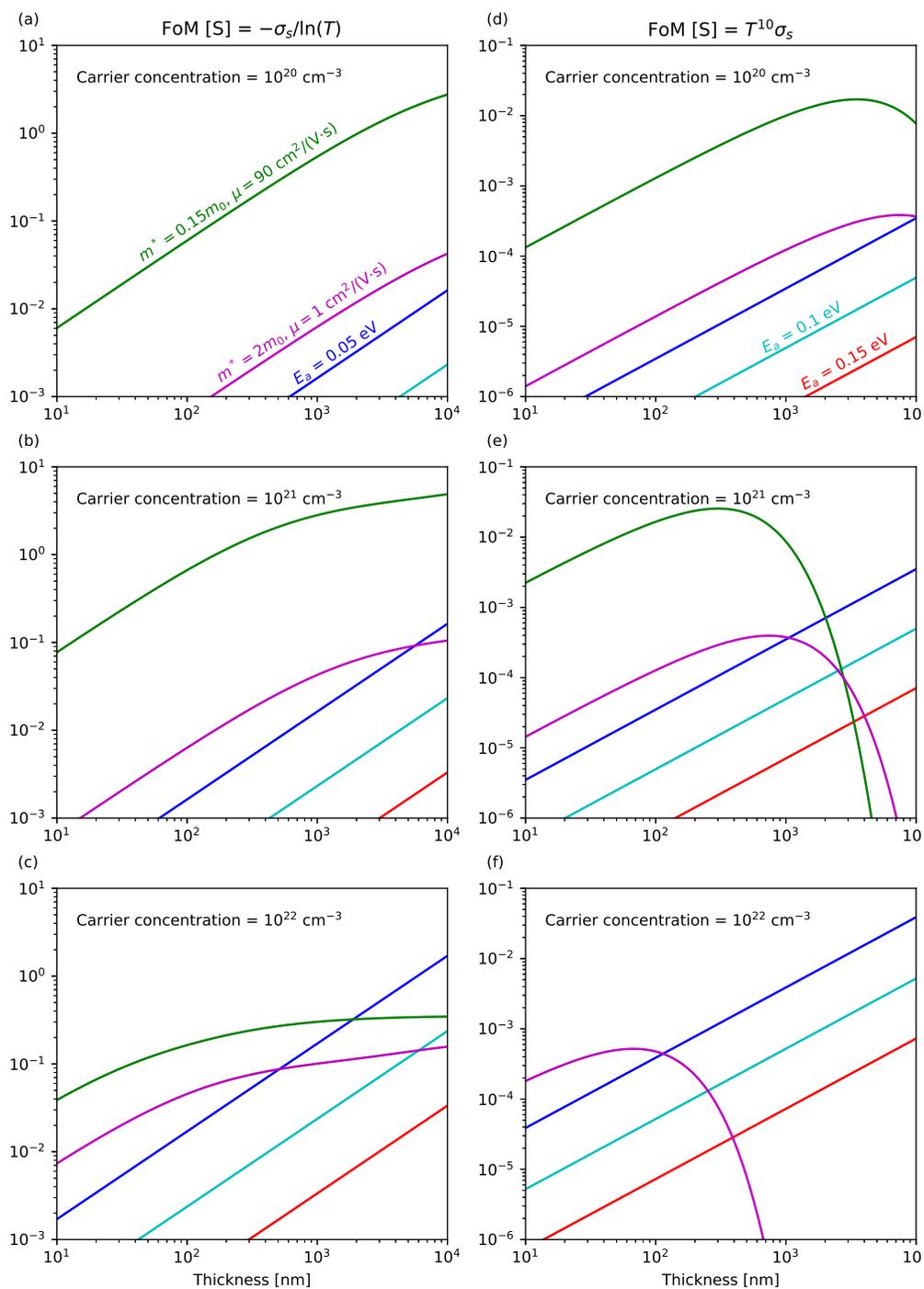}
	\caption{Gordon (left) and Haacke (right) figures of merit for different carrier concentrations as a function of the thickness of the thin film. B-TCOs with parameters typical of n- and p-type (green and purple respectively) materials are represented as well as SP-TCOs (blue, cyan and red) with various activation energies. When the carrier concentration increases, the SP-TCOs become more efficient than B-TCOs for a large enough thickness.}
	\label{fig:FoM_t}
	\end{figure}

Different definitions of the figure of merit exist in the literature. 
The one we consider depends on the targeted application : the relative weights of the conductivity and the transparency are different in each case. For example, a TCO used for IR-reflecting windows must have a high transparency, but their conductivity is not important.\cite{gordon2000criteria} The electrode of a solar cell needs to be both transparent and conducting.\cite{gordon2000criteria} Gordon's FoM is defined as $-\sigma_s / \ln(T)$.\cite{gordon2000criteria} This FoM gives a more important weight to the conductivity : materials with a transmittance as low as 30\% can have large Gordon's FoM. Its important use in literature is due to its independence on the thickness of the TCO, for a given light wavelength. Haacke's FoM ($\sigma_s T^{10}$) does not give more weight to the conductivity, and the relative weights can be tuned by modifying the exponent of the transmittance.
We did not limit our analysis to Haacke's FoM. Figure \ref{fig:FoM_t} compares both Gordon and Haacke's FoM as functions of the thickness of the TCO film for different carrier concentration. The analysis of the right side of the figure is realized in the main body of the paper. When the thickness increases, we observe that Gordon's FoM does not decrease for B-TCOs even if the absorption becomes large. The behavior of SP-TCOs is similar for both FoM because of their low absorption. Because Gordon's FoM favours the conductivity, SP-TCOs are less efficient compared to B-TCOs than they are when we use Haacke's FoM. Nevertheless, in both cases, there exist a thickness above which SP-TCOs have a better FoM than high effective mass/low mobility B-TCOs. For example, polaronic TCOs with a carrier concentration around $10^{22}$~cm$^{-3}$ and an activation energy between 0.05 and 0.1~eV have larger Gordon's FoM than lousy B-TCOs of thickness larger than 500~nm. 

\FloatBarrier
\section{LaCrO$_3$}

\subsection{Formation, transition and activation energies}
We report the formation, strain and transition energies of small polarons in La$_{1-x}$Sr$_x$CrO$_3$ in Table~\ref{tab:Transition_Energy}. 
These energies are defined in Figure~\ref{fig:Energy_Polaron}.
The strain energy $E_s$ is an order of magnitude larger than the formation energy $E_p$. 
For a value of $x$, $E_s$ depends on the configuration and the location of the small polarons but globally decreases when the distance between small polarons increases. 
$E_p$ and $E_s$ should reach converged values for $x \rightarrow 0$.
These values lead to a transition energy $E_T$ between 0.56 and 0.79~eV that can explain the absorption peak observed by Zhang \textit{et al.} at $0.7-0.8$~eV.\cite{zhang2015perovskite}
The observed peak is therefore due to the transition from the small-polaron state to the delocalized state, where carriers are band-like.

The formation energy is very small for all $x$ and the associated transition energy $2E_p$ (between 0.04 and 0.19~eV) is too small to be seen on the absorption feature obtained by Zhang \textit{et al}.\cite{zhang2015perovskite}
To our knowledge, no experimental data are available to compare with the presented results for the transition to the excited small-polaron state.
In any case, the transition energies are small enough not to lead to an important absorption of visible light, which is one of the required properties to make a good SP-TCO. 

	\begin{table}[!h]
	\center
	\caption{Formation energy $E_p$, strain energy $E_s$ and transition energies $E_T$ and $2E_p$ of small polarons in different La$_{1-x}$Sr$_x$CrO$_3$ supercells. All energies are given in eV.}
	\setlength{\tabcolsep}{8pt}
	\begin{tabular}{cccccc}
	\hline
	\hline
	\rule[-0.0cm]{0mm}{0.4cm} Supercell & $x$ & $E_p$ & $E_{s}$ &  $E_T$ & $2E_p$ \\
	 \hline
	\rule[-0.0cm]{0.0mm}{0.4cm} \hspace{-2mm} 1$\times$1$\times$1 & 1/4 & 0.09 & 0.70 & 0.79 & 0.19 \\
	1$\times$1$\times$2 & 1/8 & 0.05 & 0.67 & 0.72 & 0.11 \\
	1$\times$2$\times$1 & 1/8 & 0.02 & 0.54 & 0.56 & 0.04 \\
	1$\times$2$\times$2 & 1/16 & 0.05 & 0.57 & 0.62 & 0.11 \\
	2$\times$2$\times$1 & 1/16 & 0.09 & 0.48 & 0.58 & 0.19 \\
	2$\times$2$\times$2 & 1/32 & 0.06 & 0.50 & 0.56 & 0.12  \\
	\hline
	\hline
	\end{tabular}
	\label{tab:Transition_Energy}
	\end{table}
	
The activation energy for the hopping mechanism is reported in Table~\ref{tab:Activation_Energy}. 
They are $\sim$~0.1~eV for all our computations, which is in good agreement with other theoretical work and fitting of resistivity-temperature curves.\cite{zhang2015hole} 
The formation and the activation energies have similar values, which goes against Marcus model.
We explain this discrepancy by the high concentrations of carriers in this material which induce interactions between the small polarons that Marcus theory does not consider.
However, our previous model still works with decoupled activation and transition energies (see Section S\ref{supsec:decoupled}).
We can still understand the small-polaron transport with an activation energy and the optical properties with a transition energy which are unrelated (see Figure~\ref{fig:Polaron_decoupled}).
	
	\begin{table}[!h]
	\center
	\caption{Activation energy $E_a$ of the hopping mechanism for different concentrations. All energies are given in eV.}
	\setlength{\tabcolsep}{15pt}
	\begin{tabular}{ccccc}
	\hline
	\hline
	\rule[-0.0cm]{0mm}{0.4cm} $x$ & ${E_a}^1$ & ${E_a}^2$   & $E_a^{\mathrm{Expt.}^2}$  \\
	 \hline
	\rule[-0.0cm]{0.0mm}{0.4cm} \hspace{-2mm} 0.25 & 0.10 & 0.18 & 0.08 \\
	0.125 & 0.11 & 0.11 & 0.10 \\
	0.0625 & 0.10 & \ding{53} & \ding{53} \\
	\hline
	\hline
	\end{tabular}
	\begin{flushleft}
	$^1$ This work. \\
	$^2$ Ref.~\cite{zhang2015hole}
	\end{flushleft}
	\label{tab:Activation_Energy}
	\end{table}

\subsection{Band structures}

\begin{figure}[h]
	\center
	\includegraphics[clip,trim=0.8cm 0.3cm 1.5cm 1.3cm,width=0.9\textwidth]{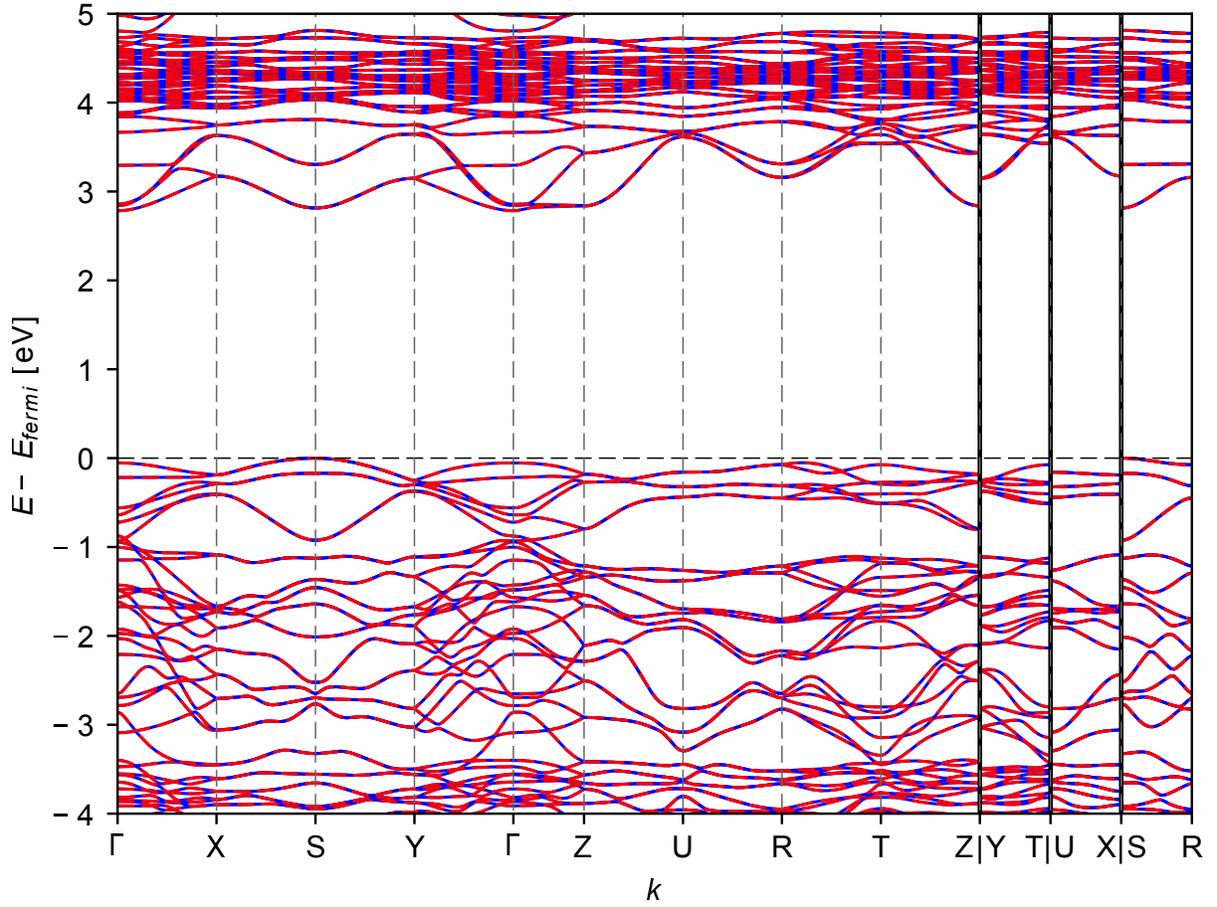}
	\caption{Band structure of LaCrO$_3$ obtained with GGA+U. Red and blue lines indicate opposite spin states. The band gap is of 2.8~eV.}
	\label{fig:LaCrO3_bs}
	\end{figure}

\begin{figure}[h]
	\center
	\includegraphics[clip,trim=0.8cm 0.3cm 1.5cm 1.3cm,width=0.9\textwidth]{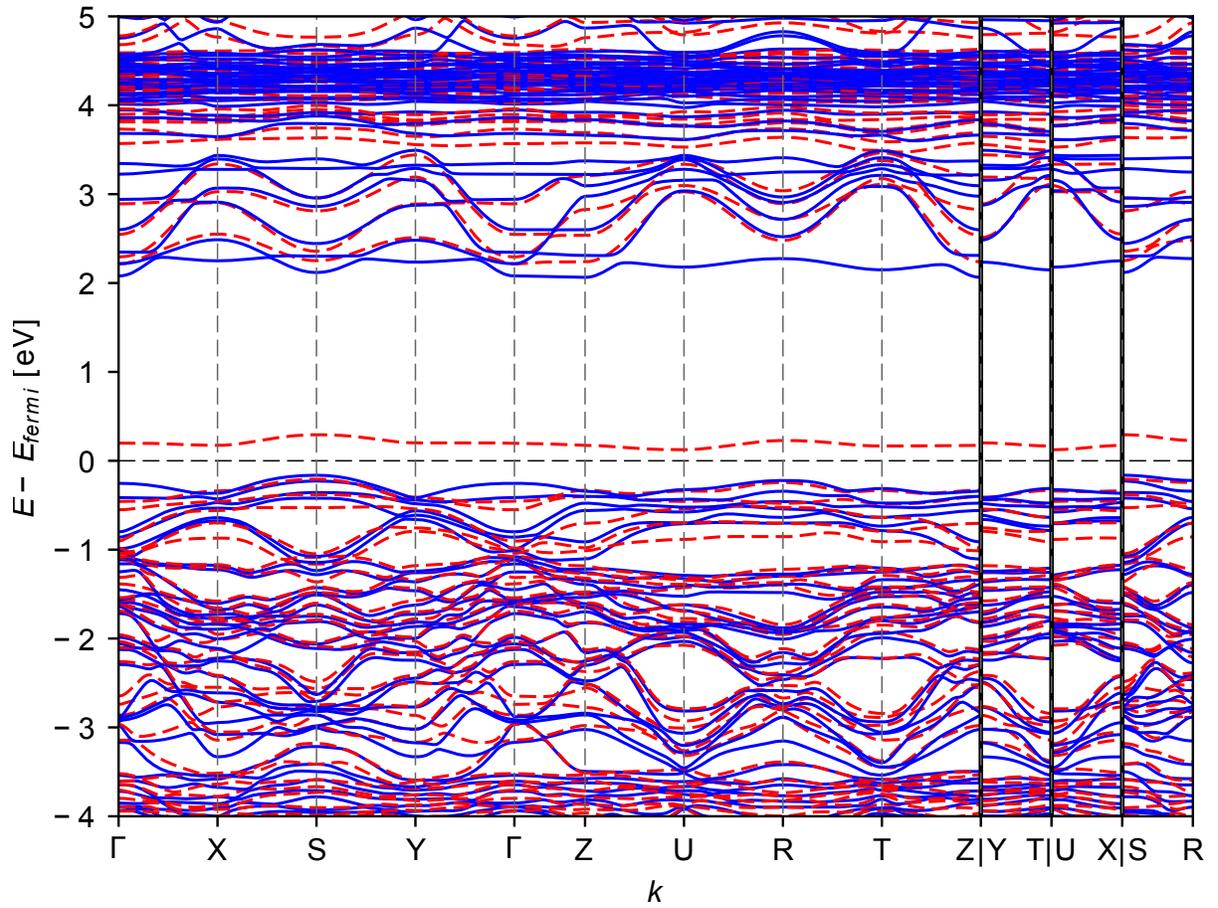}
	\caption{Band structure of Sr$_{1/4}$La$_{3/4}$CrO$_3$ obtained with GGA+U. Red and blue lines indicate opposite spin states. A flat small-polaron state appears in the original band gap which decreases to 2.2~eV.}
	\label{fig:SrLaCrO3_bs}
	\end{figure}

\begin{figure}[h]
	\center
	\includegraphics[clip,trim=1.2cm 0.6cm 1.7cm 0.9cm,width=0.9\textwidth]{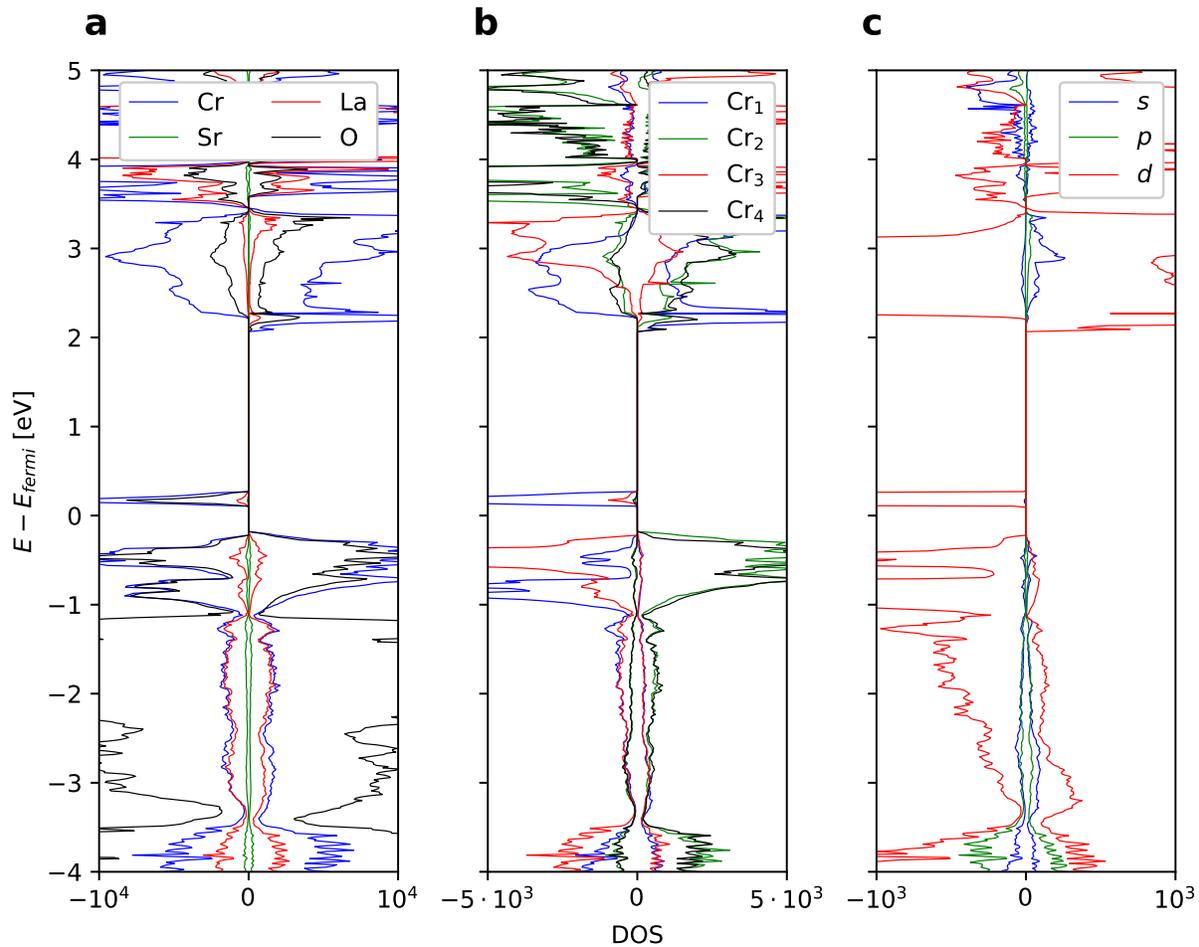}
	\caption{Projected density of states (pDOS) of Sr$_{1/4}$La$_{3/4}$CrO$_3$ obtained with GGA+U. (a) DOS projected on the different atoms of the unit cell. The localized state in the forbidden band is mainly composed of Cr with some O states. (b) DOS projected on the four Cr atoms of the unit cell. The localized state is localized on a specific Cr atom. (c) DOS projected on the orbitals of the specific Cr atom. It is a Cr$_1$ $d$ state.}
	\label{fig:SrLaCrO3_dos}
	\end{figure}

Figure~\ref{fig:LaCrO3_bs} and \ref{fig:SrLaCrO3_bs} plot band structures of LaCrO$_3$ and Sr$_{1/4}$La$_{3/4}$CrO$_3$ obtained with GGA+U. Red and blue lines correspond to opposite spin states. The band gap of LaCrO$_3$ is 2.8~eV and decreases to 2.2~eV when an atom of Sr replaces one of the four La atoms in the unit cell. A flat small-polaron state appears in the original band gap. The projected density of states is represented in Figure~\ref{fig:SrLaCrO3_dos}. The localized state formed by the apparition of the small polaron is constituted of $d$ states of one of the four Cr atoms in the unit cell.

\FloatBarrier
\subsection{Mixing energy}
The substitution of La atoms by Sr ones in LaCrO$_3$ is really easy and important doping levels can be reached. For instance, Zhang \textit{et al.} have grown samples of Sr$_x$La$_{1-x}$CrO$_3$ with $x$ in the whole range between $0$ and $1$.\cite{zhang2015perovskite,zhang2015hole} We computed the mixing energy with GGA+U in the case of $x=0.25$ and obtained 80 meV for one Sr atom in one unit cell (4 formula units). We can compare this value to the ideal entropy of mixing $\Delta S_\mathrm{mixing}^\mathrm{ideal}$ given by:\cite{chevrier2013first}
\begin{equation}
\Delta S_\mathrm{mixing}^\mathrm{ideal} = -k_B \left( \frac{1}{4} \ln \frac{1}{4} + \frac{3}{4} \ln \frac{3}{4} \right) \approx 0.048~\frac{\mathrm{meV}}{\mathrm{Sr~K}}
\end{equation}
which, at the growth temperature of 1000 K,\cite{zhang2015perovskite} is of 48 meV. This is only an estimate because we considered an ideal mixing where the species do not interact. The mixing energy is therefore on the scale of entropic contributions.

\FloatBarrier
\subsection{NEB calculations}

\begin{figure}[h]
	\center
	{\includegraphics[width=0.49\columnwidth]{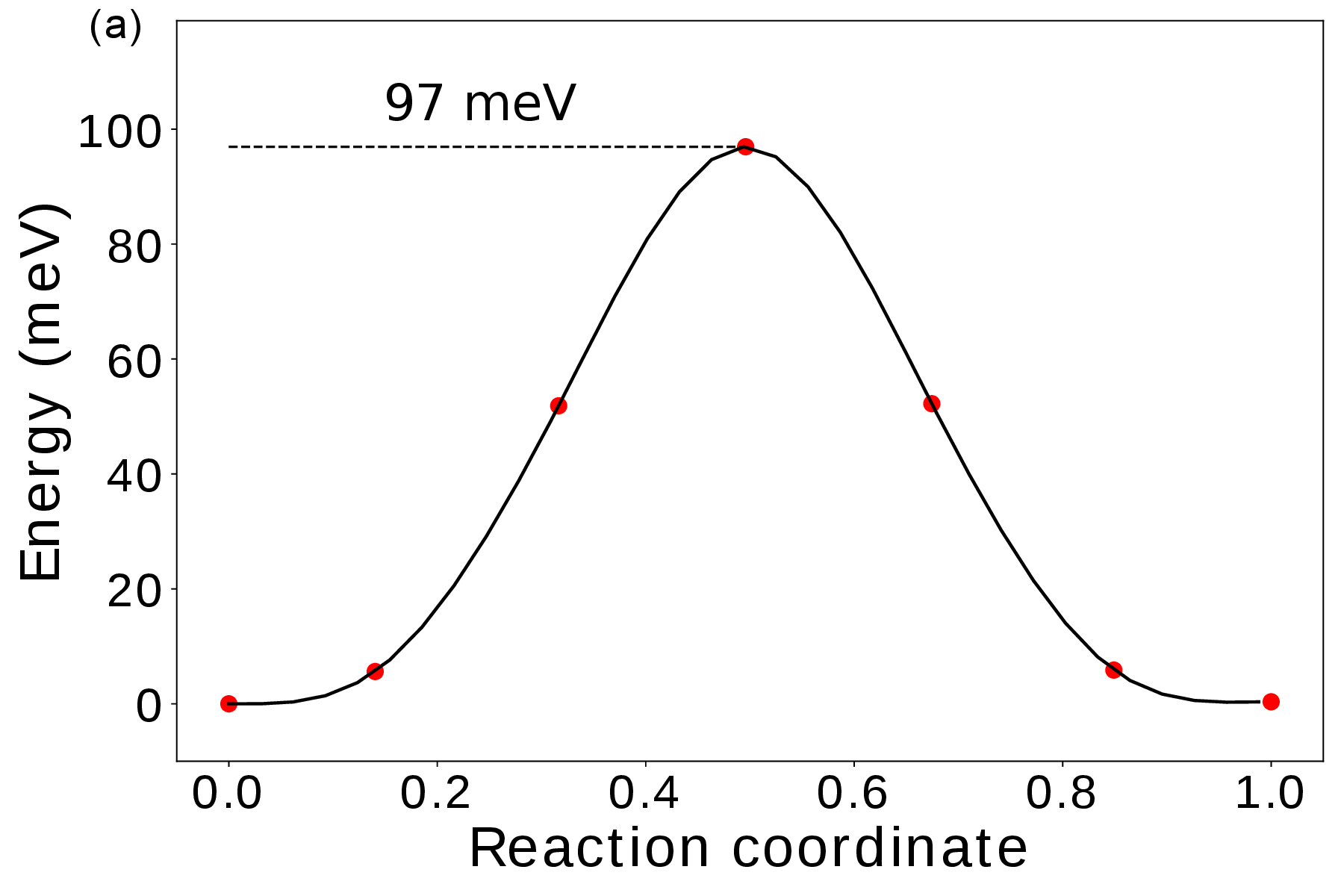}}
	
	\vspace{5mm}
	{\includegraphics[width=0.49\columnwidth]{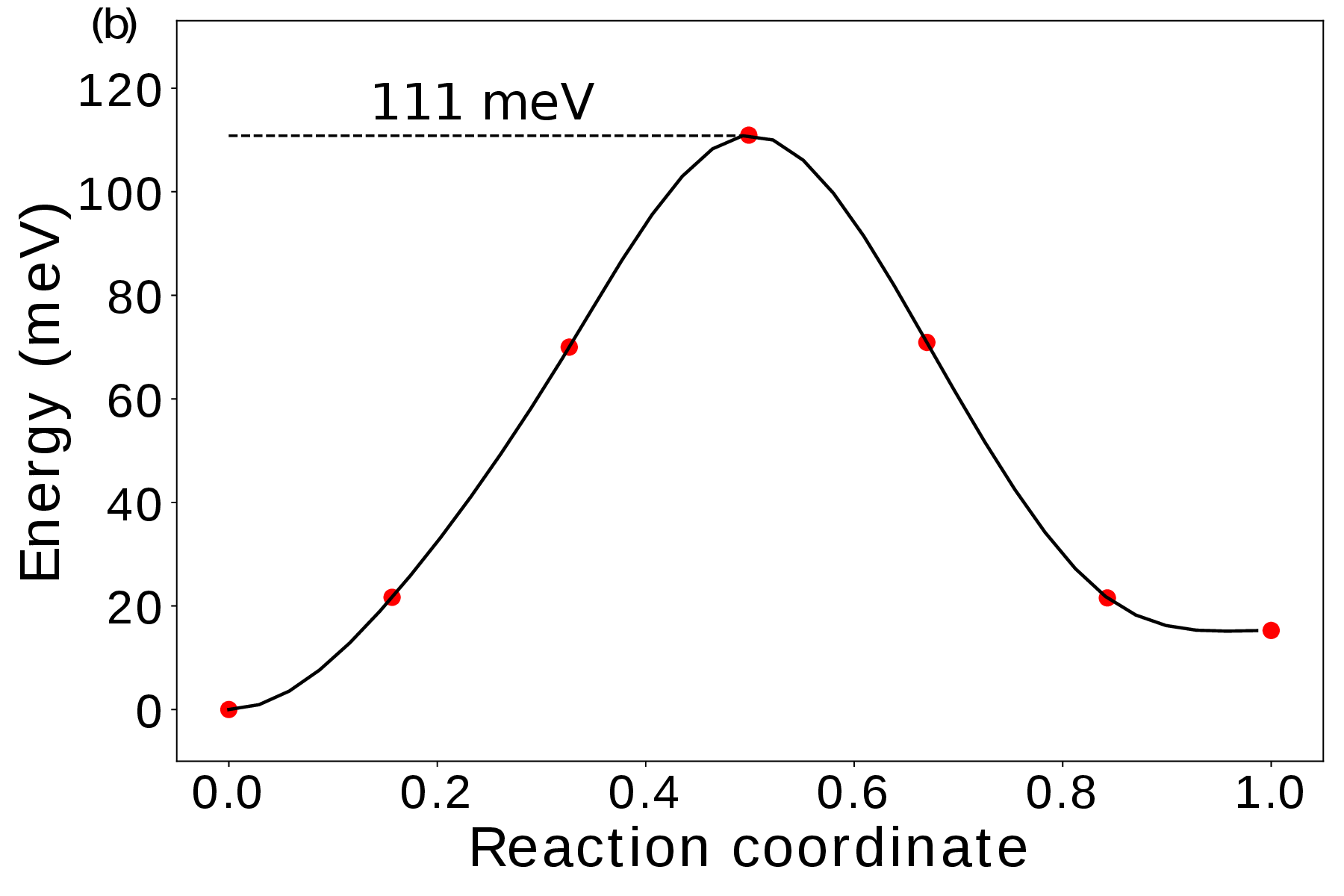}}
	
	\vspace{5mm}
	{\includegraphics[width=0.49\columnwidth]{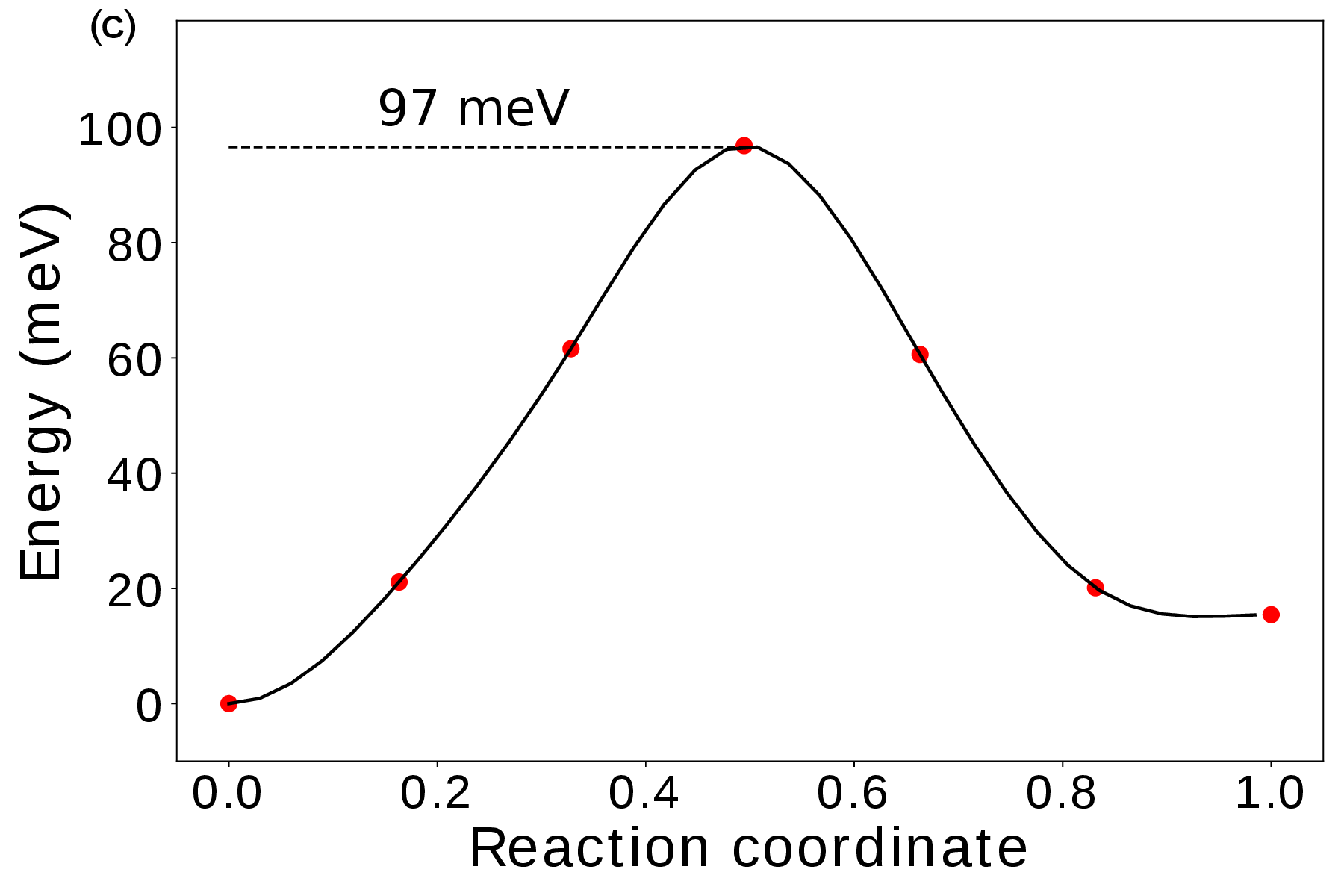}}
	\caption{Hole transfer energies a) along the [010] direction in the 1$\times$2$\times$1 LaCrO$_3$ supercell containing two holes b) along the [010] direction in the 1$\times$2$\times$1 LaCrO$_3$ supercell containing one hole and c) along the [100] direction in the 2$\times$2$\times$1 LaCrO$_3$ supercell.}
	\label{fig:Polaron_Transfer_Path}
	\end{figure}

Figure~\ref{fig:Polaron_Transfer_Path} gives the small-polaron transfer paths along a single direction in different LaCrO$_3$ supercells. In each case, the activation energy is $\sim$~0.1~eV which is in good comparison with the experimental value.\cite{zhang2015hole}

\renewcommand{\refname}{Supplementary References}

\bibliographystyle{ieeetr}
\bibliography{References}